\documentclass[preprint]{aastex}
\usepackage{epsfig}
\usepackage{natbib}

\newcommand{\bmaxr}{t(B$_{max})$}
\newcommand{\rmax}{T$(r_{max})$}

\def\gsim{\;\rlap{\lower 2.5pt
 \hbox{$\sim$}}\raise 1.5pt\hbox{$>$}\;}
\def\lsim{\;\rlap{\lower 2.5pt
   \hbox{$\sim$}}\raise 1.5pt\hbox{$<$}\;}

\begin{document}

\title{Photometric Identification of Type Ia Supernovae at Moderate Redshift}
\author{Benjamin D. Johnson and Arlin P. S. Crotts}
\affil{Department of Astronomy, Columbia University,  550 West 120th St, New York, NY 10027, USA}
\email{bjohnson@astro.columbia.edu; arlin@astro.columbia.edu}


\begin{abstract}
Large photometric surveys with the aim of identifying many Type Ia supernovae (SNe) at moderate redshift are challenged in separating these SNe from other SN types. We are motivated to identify Type Ia SNe based only on broadband photometric information, since spectroscopic determination of the SN type, the traditional method, requires significant amounts of time on large telescopes.  We consider the possible observables provided by a large synoptic photometry survey. We examine the optical colors and magnitudes of many SN types from $z=0.1$ to $z=1.0$, using space-based ultraviolet spectra and ground-based optical spectra to simulate the photometry.  We also discuss the evolution of colors over the SN outburst and the use of host galaxy characteristics to aid in the identification of Type Ia SNe.  We consider magnitudes in both the SDSS photometric system and in a proposed filter system with logarithmically spaced bandpasses.  We find that photometric information in four bands covering the entire optical spectrum appears capable of providing identification of Type Ia SNe based on their colors at a single observed epoch soon after maximum light, even without independent estimates of the SN redshift.  Very blue filters are extremely helpful, as at moderate redshift they sample the restframe ultraviolet spectrum where the SN types are very different. We emphasize the need for further observations of SNe in the restframe ultraviolet to fully characterize, refine, and improve this method of SN type identification.
\end{abstract}

\section{Introduction}
Observations of Type Ia supernovae (SNe) have become a cornerstone of physical cosmology, providing some of the first hints of the presence of a cosmological constant, or dark energy \citep{reiss98,perlmutter99}.  Efforts to increase the robustness and precision of this result using Type Ia SNe fall into two categories.  The first is the observation of Type Ia SNe at redshifts $z$ greater than 1, where the effect of a cosmological constant becomes negligible and the luminosity distance~-~redshift relation for models containing a cosmological constant or dark energy deviates from most alternative models involving, for example, grey dust \citep{aguirre99}.  A second method for verifying the SN results involves using large numbers of moderate-redshift ($z \lsim 1$) Type Ia SNe to investigate systematic effects such as dust extinction and SN evolution by comparing subsamples \citep{branch_sub}. Such a sample of moderate-redshift Type Ia SNe may also be used to constrain the equation of state of the dark energy. A combination of both methods will aid in constraining the time variation of the dark energy density \citep{wang01}, leading to constraints on the source of dark energy.  It is thus interesting to observe both some high redshift and many moderate-redshift Type Ia SNe.

For photometric surveys designed to discover large numbers of moderate-redshift SNe, it is not practical to determine the type of every SN via spectroscopy. Additionally, if the SNe are not discovered in nearly real time then spectroscopic followup may be impossible.  Another way to identify the cosmologically useful Type Ia SN is therefore necessary. We examine here the idea of photometric identification of Type Ia SNe in the context of large synoptic photometric surveys.  Aspects of such photometric identification have been explored extensively in \citet{poznanski}. \citet{Reiss03} provide an application of this method to very high redshift ($z>1.5$) SNe discovered with HST.  Here we combine and extend these two studies by examining the blue magnitudes of moderate-redshift SNe ($z \le 1.0$), making use of the restframe ultraviolet differences between SN types that was noted by \citet{Reiss03}.  In a shift of emphasis from \citet{poznanski} we explore the use of photometric properties for SN type identification when independent redshift information is \emph{not} available but when temporal information is available. We explore the possibility that identification of Type Ia SNe can be accomplished using information at a single observed epoch in the SN lightcurve, provided that epoch is known.  For the first time we attempt to estimate the uncertainties in this method due to the dispersion in observed SN colors that may arise from many sources.  We also discuss the use of other information that is likely to be available in typical synoptic surveys.

Proposed projects such as the LSST \footnote{http://www.noao.edu/lsst}, which would observe the entire visible sky once every few days, may benefit from a photometric method of typing SNe.  By photometrically identifying objects which are unlikely to be Type Ia SNe the LSST or other large, rapid-cadence, SN surveys could greatly increase the efficiency of campaigns to observe Type Ia SNe spectroscopically.  It may be possible that for certain restricted studies the spectra will not be necessary at all \citep{barris}.  Most other objects that vary with the timescale and amplitude of SNe have very different colors (e.g. Cepheids), are repeating, or are very rare (e.g. long-duration novae).  For this reason we concentrate on separating Type Ia SNe only from other SN types.

\section{Photometric SN Observables}
  The aim of this study is to determine which set of observables is most advantageous for separating Type Ia SNe from other types, including which filters are most useful or necessary.  We assume here that the SNe will be detectable in several wavelength bands with adequate signal-to-noise (S/N) for at least 10 rest-frame days after maximum (for SNe Ia this translates into $\lsim$ 1 mag. below maximum light).  

Thus the magnitude, several colors, and temporal information will be available to aid in the identification/separation of Type Ia SNe. Note that color, magnitude, and morphologoical information may also be available for the SN host galaxy. In \S~\ref{sec:color} we begin the discussion of type separation by using broadband colors at a single observed epoch to discriminate between SNe.  In \S~\ref{sec:mag} we discuss the addition of the magnitude observable to the color observable, and in \S~\ref{sec:time} we discuss combinations of observables that include time as well (i.e. multiple epochs).  In each case we examine selected subsets of the observables. We are prevented from considering all observables at all redshifts simultaneously due to insufficient spectral data. Throughout the discussion in these sections it is assumed that independent redshift information is \emph{not} available. In \S~\ref{sec:host} we discuss the possibility that some information about the host galaxy will be available as well, especially for the lower redshift sources.  This may include the morphology and/or a photometric redshift.

\begin{figure}
\plotone{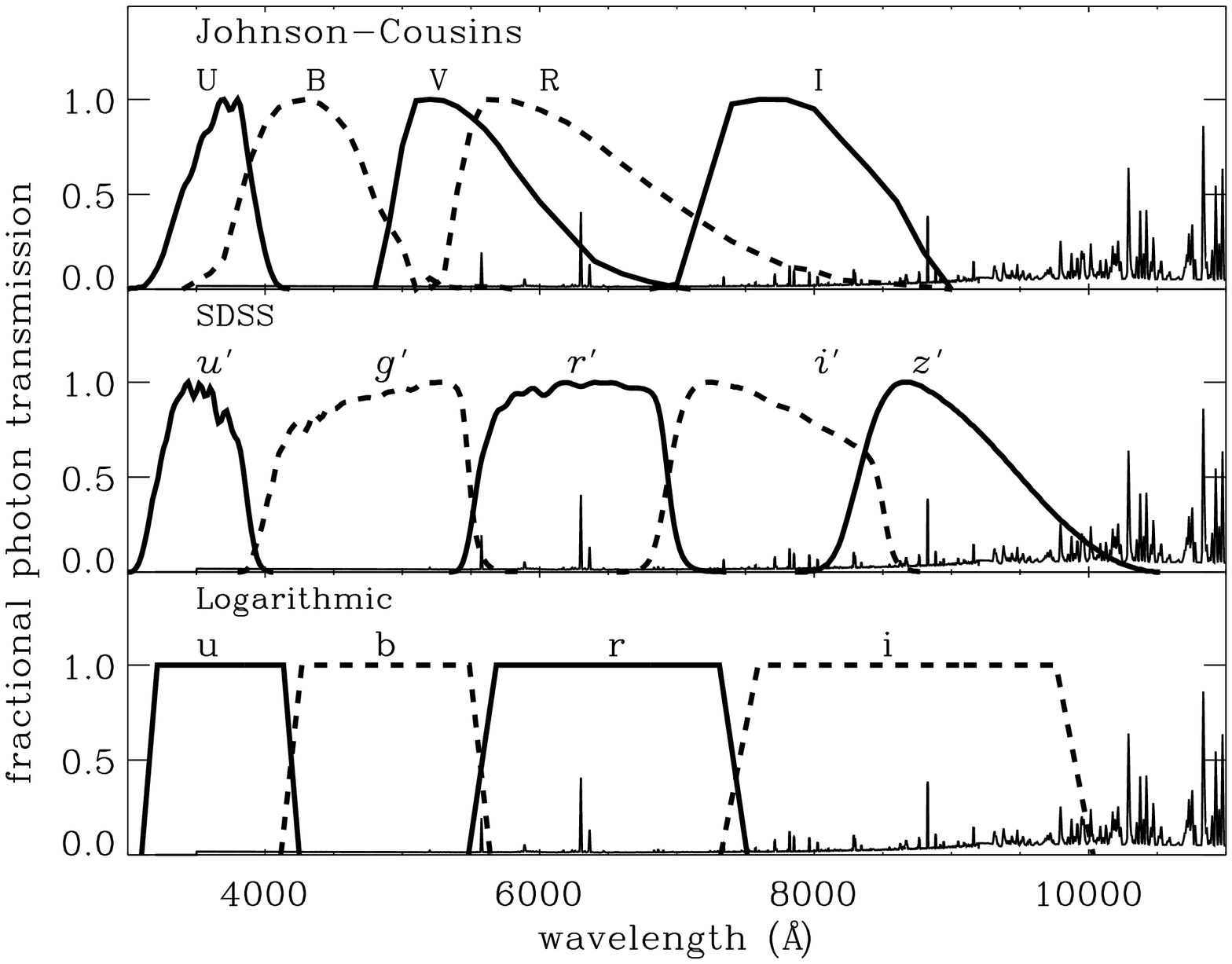}
\figcaption{\emph{Top}: The \citet{Bessel90} standard Johnson/Kron-Cousins \emph{photon} transmission functions overlayed on a Mauna Kea dark sky spectrum.\emph{Middle}: The SDSS fliter transmission functions.\emph{Bottom}: A hypothetical logarithmically spaced filter system.  The cut-on and cut-off of these filters have been altered from simple step functions to reduce the effect of narrow spectral lines on the photometry, as would be true of most real filter systems.\label{fig:filters}}
\end{figure}

In the following we denote Johnson-Cousins filters with capital letters, and the SDSS filters with primed lower case letters. We also investigate the use of a filter system in which the central wavelengths are logarithmically spaced (hereafter referred to as the logarithmic system). This filter system has been designed such that the entire optical region is divided into 4 filters that redshift into each other (at redshift intervals of $(1+z)\approx 1.35^{n}$), minimizing the cross-filter k-corrections at these redshifts.  It shares with the SDSS system the advantage that the OI$\lambda \lambda 5507$ night sky emission falls between 2 of the filters. These filters will be labeled with lowercase unprimed letters.  These three filter systems are shown in Figure \ref{fig:filters}.  All synthetic magnitudes and colors are on the AB system \citep{oke}. All epochs $\tau$ are with respect to the quoted date of SN restframe maximum light in the B band, \bmaxr, and are given in days in the SN rest frame unless otherwise noted.    Throughout this study we use a concordance cosmology: $\Omega_{m}=0.3,   \Omega_{\Lambda}=0.7,  \mbox{H}_{\circ}=70$ km s$^{-1}$ Mpc$^{-1}$.

\section{The Template Spectra}
We investigate the appearance of SNe in the photometry data using synthetic magnitudes derived from template spectra which are described below. We have two classes of spectra: a few with extensive UV coverage from space telescopes, and many with poor UV coverage from ground-based telescopes.

\subsection{UV+Optical Spectra}
\begin{figure}
\plotone{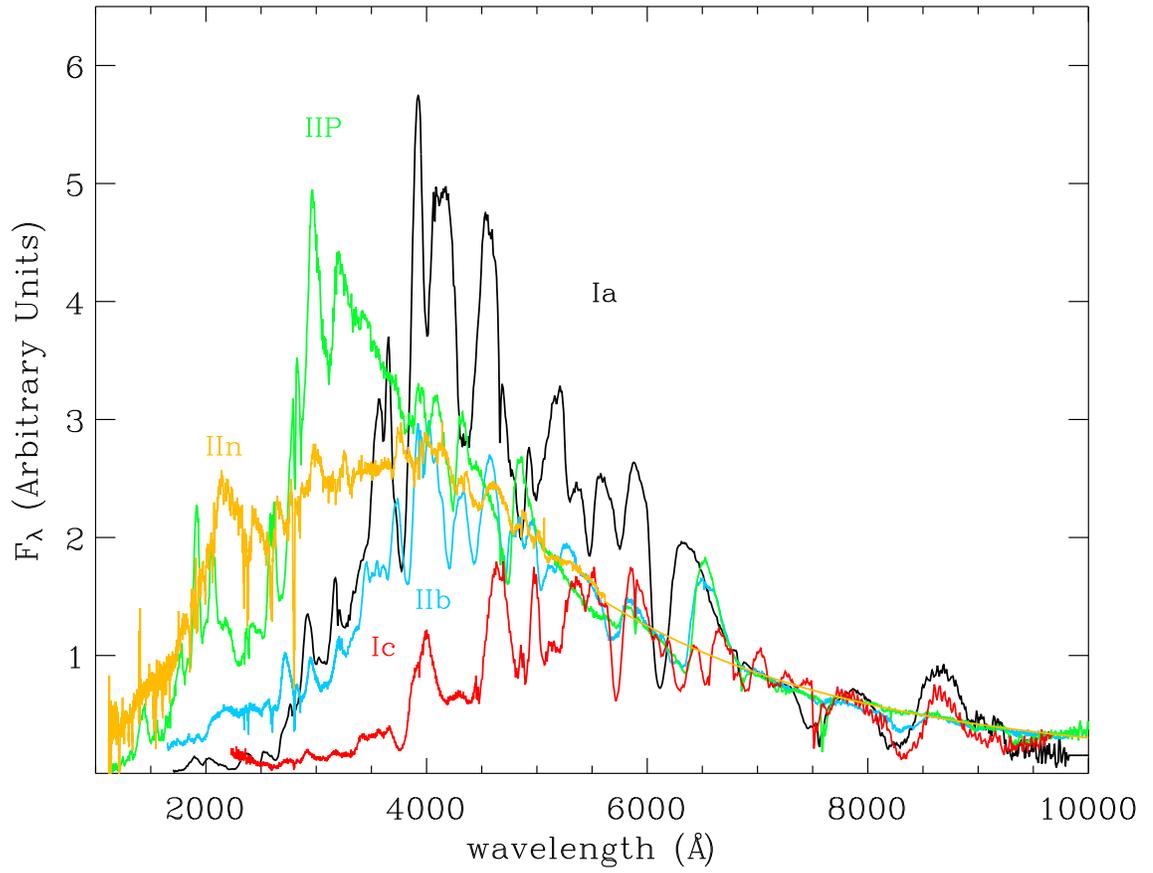}
\figcaption{The template UV+O spectra, normalized at $\sim 7200\AA$.  The Type Ia spectrum is from $\tau=5$.\label{fig:spectra}}
\end{figure}

Table \ref{table:uvo} gives the details of ultraviolet+optical (UV+O) template spectra with extensive UV coverage, and they are shown in Figure \ref{fig:spectra}. Every major SN type except Type IIL\footnote{Type IIL SNe have only been observed in the UV by IUE, and there is a significant gap in the wavelength coverage between the IUE spectra and the available optical spectra, making normalization difficult at best.} is represented.   We combine Type Ib and Type Ic into the single class Type Ib/c since at the epochs considered here they present very similar broadband colors \citep{poznanski}.   The UV spectra are taken from the preview data generated by the Multimission Archive at Space Telescope (MAST)\footnote{http://archive.stsci.edu/data.html}.  All the UV spectra are from HST STIS or FOS, except for the SN1992A $\tau=9$ days spectrum, which is from IUE.  Except for the hybrid spectra discussed below, there was significant overlap betweeen the optical and the UV wavelength coverage, and the UV spectra were well matched to the optical spectra after multiplication by a constant (with a value on the order of 1). These UV+O spectra have been deredshifted and dereddened according to the values given in Table \ref{table:uvo} (see the references for a discussion of the reddening values).  The extinction law used is that of \citet*{CCM89} as modified by \citet{odonnell94}, with R$_{V}=3.1$. We note that in many cases the values of A$_{V}$ are highly uncertain as there is no conclusive way to quantify the SN host extinction and reddening, especially towards core-collapse SNe. 

The choice of $\tau$ for these templates was determined entirely by the availability of good quality UV spectra within 20 days of \bmaxr.  This reflects the consideration that spectroscopic followup of Type Ia SN candidates should be conducted as close to maximum light as possible. We note that the Type IIb SN template spectrum is from $\tau=+0$ days, in contrast with the other template spectra, and that the reddening to SN1993J has not been well determined.

\subsection{Justification of Hybrid Spectra}
Two of the UV+O templates were created by combining the UV spectrum of a supernova with other sources.  This was necessitated by the lack of publicly available digital optical spectra at the epoch of the UV spectrum.  We here justify this method for creating the template spectra in question.

\subsubsection{Type Ib/c: SN1994I and SN1999ex}
\label{sec:specIc}
SN 1994I was observed by HST FOS on 1994 April 18 UT, and extensive optical spectroscopy of this SN has been carried out.  However, the only spectrum which is publicly available in machine-readable form is from 1994 April 8 UT. Optical spectra of SN1999ex at $\tau=13$ days, on the other hand, have been made available in digital form. Both SNe were Type Ic and displayed very similar spectra \citep{IcSpec}.  It is the strength of this similarity that gives us the confidence to form this hybrid spectrum.  Unfortunately, the color evolution of these two SNe differed significantly.  \citet{max} show that the B$-$V color of SN1994I was 0.25 mag redder than SN1999ex at $\tau=$10.  This is due to the fast evolution of SN1994I, which resembles that of SN1999ex compressed in time.  This is a general problem when exploring the colors of Type Ib/c SNe that is discussed in more detail below.

\subsubsection{Type IIn: SN1998S and a Blackbody}
SN1998S displayed a relatively featureless blue continuum during its early evolution ($-10 < \tau < 40$ days).  \citet{leonard2000} show that the March 27th spectrum ($\tau \approx 7$, deredddened with E(B$-$V)=0.1) can be approximated by a T=10000$^{\circ}$K blackbody to within 10\% for $\lambda>5000$ \AA. \citet{fassia_spec} show a March 30 spectrum with similar shape. We therefore approximate the $\tau = 10$ unreddened spectrum with a blackbody curve for $\lambda>5500$ \AA.  As we mention below the V$-$R and R$-$I colors of this template agree well with the published photometry, thus increasing our confidence in the blackbody approximation of the spectrum.

\subsection{Other Spectra}
We have also collected many publicly available SN spectra, mostly from the SUSPECT database\footnote{http://bruford.nhn.ou.edu/\~ suspect}, which do not have good UV coverage (Table \ref{table:speclist}).  They have been deredshifted and dereddened according to the values given in Table \ref{table:speclist}, and with the same reddening law described above. We will use these spectra to estimate the homogeneity (or lack thereof) of the colors of various SNe types near the UV+O template epoch.  Also, these spectra span a range of epochs and will be used in the temporal analysis of spectra in \S~\ref{sec:time} , i.e. in the examination of the color evolution of SNe as a possible SN type discriminator.   Studies involving these spectra with poor UV coverage are necessarily limited to low redshift and/or the redder filters.

\section{Color Separation: Color-Color Diagrams}
\label{sec:color}
We first attempt to separate Type Ia SNe from other SN types over a range of redshifts on the basis of their colors at a single \emph{observed} epoch with respect to the \emph{observed} date of maximum light, since we are assuming here that the redshift is not independently known.  This is likely to be the problem faced by many synoptic SN surveys. Note that we implicitly assume here that enough lightcurve (i.e time and magnitude) information is available to identify the date of maximum light in a given observed band.  Because of the paucity of existing UV SN spectra we are forced to consider only one observed epoch when investigating the appearance of SNe in the bluer bands and at higher redshifts.  True multicolor SN surveys are likely to have measurements in several bands over a larger range of epoch.  We address this in \S~\ref{sec:time} below for the filter and redshift combinations where this is possible. Here we consider only one observed epoch, assuming that enough measurements will exist in at least one band to identify the time of maximum light in that band to reasonable accuracy ($\pm$ 1 day).

In attempting to compare colors at a single observed epoch with respect to observed maximum light we must account for two systematic effects.  First, for a given filter the time of SN maximum light in that filter (with respect to \bmaxr) changes with redshift. This is because redshifting causes different portions of the restframe spectrum to be sampled by any one filter and different parts of the spectrum may peak at different times.  The general trend of this effect for all SN types is that the bluer parts of the spectrum peak before the redder parts, since the photosphere cools as the SN evolves.  Second, cosmological time dilation will also change the relationship between $\tau$ and the observed epoch. 

It is important to have clear notation when relating the observed epochs to $\tau$.  We write the date of \emph{observed} maximum light in band \emph{f} for a Type Ia SN, regardless of redshift, as T$(f_{max})$. We write the band \emph{f} blueshifted to the redshift of the SN as $^{z}f$ \citep{nyu_kcorr}. We write the date of \emph{restframe} maximum light in band $^{z}f$ as t$(^{z}f_{max})$. Then we can write
\begin{equation}
\tau = [\mbox{t}(^{z}f_{max})-\mbox{t(B}_{max})]+\frac{\mbox{UT}_{obs}-\mbox{T}(f_{max})}{(1+z)}
\end{equation}
where UT$_{obs}$ is the date of the observation (in days) and $z$ is the redshift of the SN.  The first term accounts for the change in $\tau$ due to the different wavelength region being observed and the second term accounts for the change in $\tau$ due to the time of observation and cosmological time dilation. Both terms are dependent on the redshift of the SN. In what follows the distinction between T (observed frame) and t (restframe) will be important.

Due to the availability of spectral data we choose $\mbox{UT}_{obs}$=T($r_{max})+8$ observer frame days as the fiducial \emph{observed} epoch at which we compare the SN colors.  As shown below this results in restframe epochs $\tau$ which are close to those of the template spectra for a wide range of redshifts.  With more complete template spectra we would be able to compare colors at several observed epochs.  We address this problem in \S~\ref{sec:time} for a limited redshift and filter set that does not sample the restframe $\lambda < 3600$ \AA spectrum of SNe.

As an example we consider a Type Ia SN at $z=0.1$ and at $z=0.5$.  The blueshifted band $^{0.1}r$ is approximately equal to the R band, and since t$(\mbox{R}_{max})-$\bmaxr$=2\pm 1$ restframe days (see the template lightcurves of \citet{leibundgut91}) then t$(^{0.1}r_{max})-$\bmaxr$=2\pm 1$ restframe days as well.  Thus $\mbox{UT}_{obs}$=T($r_{max})+8$ observer frame days corresponds to $\tau \approx 9$ restframe days.  Similary, for a Type Ia SN at $z=0.5$ the blueshifted band $^{0.5}r$ is approximately equivalent to the B band, and so t$(^{0.5}r_{max})-$\bmaxr$=0\pm 1$ days and $\mbox{UT}_{obs}$=T($r_{max})+8$  corresponds to $\tau \approx 5$ days.  Thus template spectra from different epochs $\tau$ should be used when calculating the colors of a SN at a given \emph{observed} epoch.  For the Type Ia colors we can do this by using the $\tau=9$ SN Ia UV+O spectrum to calculate colors at $z=0.1$ and the $\tau=5$ SN Ia UV+O spectrum to calculate colors at $z=0.5$. We show the colors for both spectra in all subsequent figures.  Given the similarity of the $r$ and $r'$ filters, the results are not significantly changed if we choose $\mbox{UT}_{obs}$=T($r'_{max})+8$ observer frame days as the fiducial.

Similar considerations can be made for Type Ib/c and Type II SNe (though for Type IIP SNe  maximum light may not be well defined, especially in the redder bands), with similar results (T($r_{max})+8$ corresponds to $\tau \sim 4-16$ days).  However, for these SN types we have only one template UV+O spectrum, so the colors cannot be adjusted for this effect.  For this reason we will add the systematic color differences caused by this effect to the error estimate below, treating our lack of multi-epoch spectral templates as, essentially, an uncertainty in the restframe epoch of the SN.  

\begin{figure}
\epsscale{0.5}
\plotone{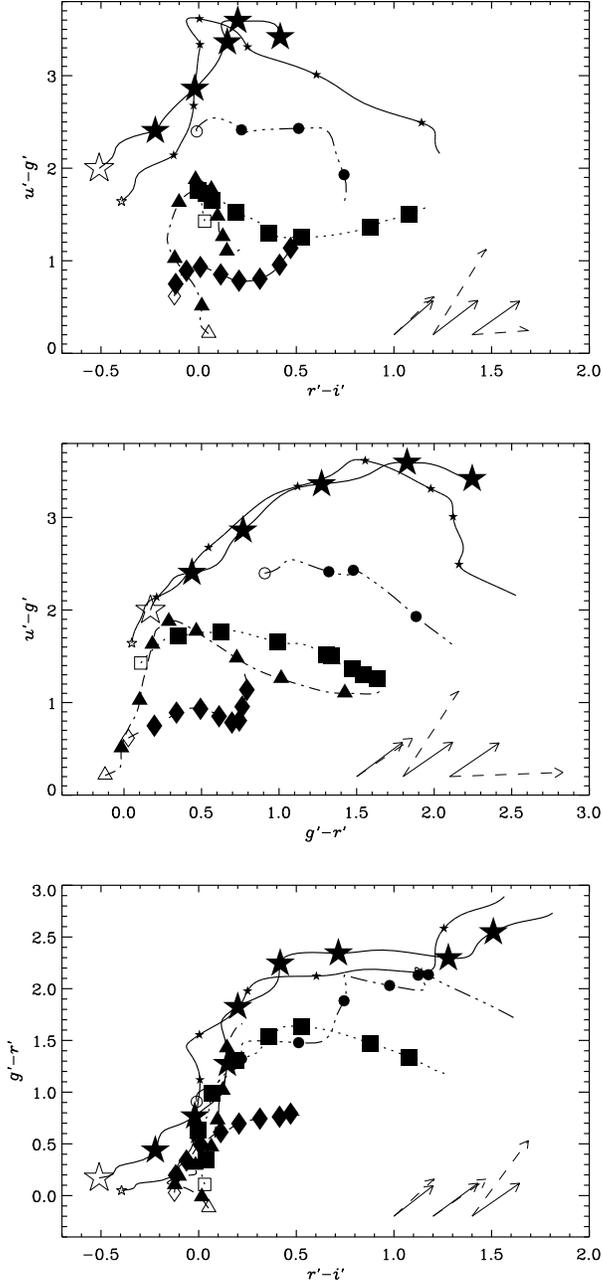}
\figcaption{SN colors as a function of redshift for the SDSS filter system.  The symbols encode the different SN types: large star,solid~-~Ia,$\tau=9$; small star,solid~-~Ia,$\tau=5$; circle,dash dot dot dot~-~Ib/c; triangle,dash dot~-~IIP; square, dotted~-~IIb; diamond, dashed~-~ IIn. Symbols are placed at intervals $\Delta z =0.1$, beginning with $z=0.1$ (open symbol for each type). The maximum redshift is determined by the spectral coverage of the template.  Arrows show, from the left to right, the effect of A$_{V}=1$ reddening at $z=0.1,0.5$, and $1.0.$ respectively for a \citet{CCM89} dust law (dashed) and a \citet{calzetti94} dust law (solid). {\it Top}: $u'-g'$ vs. $r'-i'$. {\it Middle}: $u'-g'$ vs. $g'-r'$. {\it Bottom}:$g'-r'$ vs. $r'-i'$. \label{fig:sdss}}
\end{figure}

\begin{figure}
\plotone{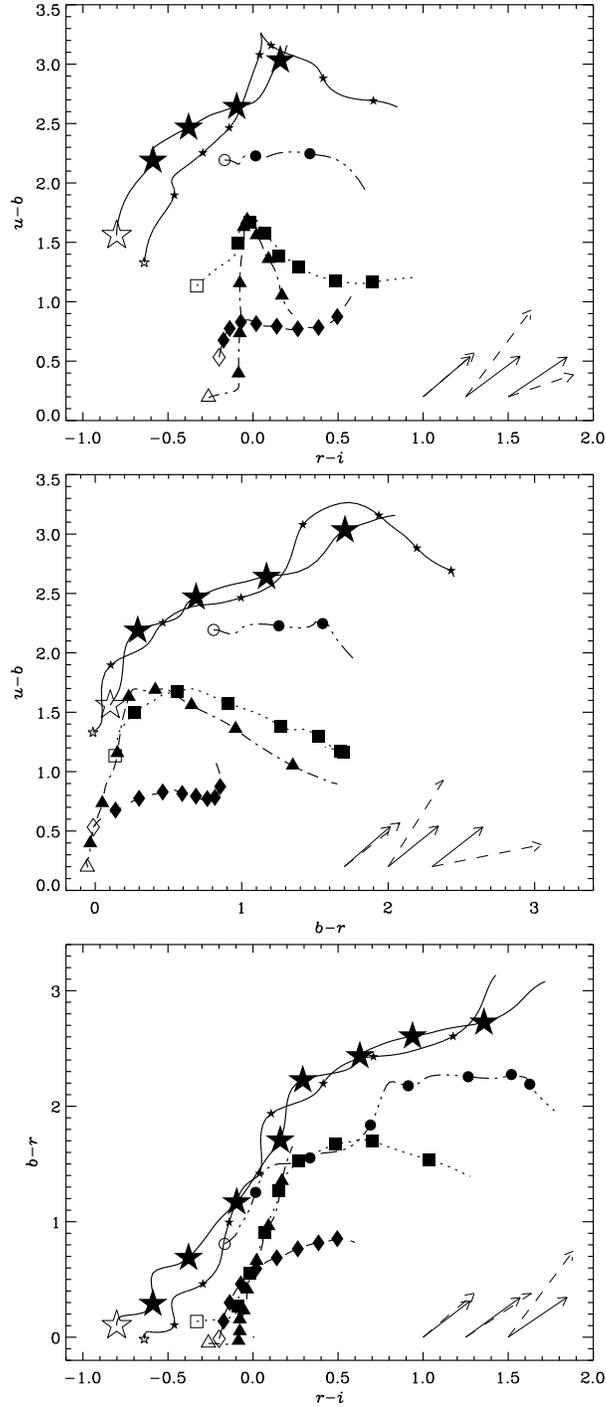}
\figcaption{SN colors as a function of redshift for the logarithmic filter system.  The symbols and arrows are as in Figure \ref{fig:sdss}. {\it Top}: $u-b$ vs. $r-i$. {\it Middle}: $u-b$ vs. $b-r$. {\it Bottom}:$b-r$ vs. $r-i$.\label{fig:log}}
\end{figure}

We use the spectra listed in Table \ref{table:uvo}, redshifted in steps of $\Delta z= 0.01$, and the filter+CCD transmission curves of Figure \ref{fig:filters} to construct synthetic magnitudes.  The results for the SDSS and logarithmic systems are shown in Figures \ref{fig:sdss} and \ref{fig:log}.  Note that for the Type Ib/c and Type Ia spectrum we do not have complete UV wavelength coverage for the bluest filters at the highest redshifts.  


We see that for both the SDSS and logarithmic filter systems the redder filters give better separation of Type Ia SNe from other types for lower redshifts, while the bluer filters give better separation at higher redshift.  Combining four filters into a $u'-g'$ vs. $r'-i'$ diagram (Figure \ref{fig:sdss}) gives the best separation over all redshift ranges for the SDSS system.  Comparable, and perhaps even better, results are obtained for the $u-b$ vs. $r-i$ colors in the logarithmic system (Figure \ref{fig:log})

\subsection{Dispersion in SN Colors}
\label{sec:disp}
For this method of identifying Type Ia SNe to be useful it is important to quantify the expected or typical difference between the colors of the different SN types calculated above from templates and the colors of a real, observed SN.  A difference between real, observed colors and those calculated from templates can have several sources, which we list in detail in Table \ref{table:errors}.  We here attempt to quantify these differences, which will be referred to as $\delta c$, for each of the SN types.

\subsubsection{All SN Types}
Several of the effects listed in Table \ref{table:errors} can be considered simultaneously for all SN types, and we address them here.  First, there are the expected photometric errors of the SN observation. These directly affect the $\delta c$  since the photometric errors can lead to inaccurate measurement of the SN color.  For S/N $> 10$  the photometric errors are typically $< 0.1$ mag and the error in color is $<$ 0.14 mag if we assume the photometric errors are uncorrelated. This error will certainly depend on redshift and SN type, but we take 0.14 mag as an upper limit

There are also errors associated with our method for creating synthetic spectra.  These errors affect $\delta c$ by giving a color for the template SN which is incorrect, and therefore not representative of the SN type as a whole.  The error due to the numerical integrations involved is less than 0.005 magnitudes, which is negligable compared to other sources.  The error due to inaccurate spectrophotometry is investigated below.  We note that only spectrophotometric errors in the template spectra which are a function of wavelength are of importance.  

To investigate possible errors in the template spectrophotometry we have calculated synthetic magnitudes on the Johnson-Cousins system, using the transmission functions of \citet{Bessel90} and the template UV+O spectra, reddened again by the values given in Table \ref{table:uvo}. The \citet{Bessel90} transmission functions have been adjusted for use with photon spectra following \citet{hamuy_specphot}. The conversion from AB to Vega based magnitudes has been determined using the \citet{kurucz} model for the spectrum of Vega, with normalization at $\lambda=5000\AA$ from \citet{HL85}.  We then compare the colors calculated in this way to the published Johnson-Cousins photometry.  We find that for the most part the \emph{colors} agree to within 0.07 mag (even for SN19998S+b-body).  The exceptions to this are some of the U$-$B colors and the colors of the Ib/c template spectrum.  The larger discrepancy of the synthetic U$-$B colors is most likely related to actual U band effective filter functions that are significantly different from our assumed filter transmission function and are therefore difficult to transform to the standard system (this is a common problem in the U band; see \citet{suntzeff_98bu} and \citet{93Jphot}). Also, the effects of atmospheric extinction are more prounounced in the U band. It is possible that there are significantly larger spectrophotometric errors in the SN templates in U band, and farther in the ultraviolet where there is no photometry to provide a check.  However, we believe this is unlikely due to the close correspondence of the UV and optical spectra in the region of overlap. This region of overlap often encompasses a significant portion of the B and V bands where the synthetic colors agree well with published photometry. However, the discrepancy between the synthetic and published colors of the template Type Ib/c spectrum is clearly due to the hybrid nature of the spectrum since the U$-$B and B$-$V colors are close to those of SN1994I but the V$-$R and R$-$I colors are close to those of SN1999ex (see \S~\ref{sec:specIc}).  

As mentioned above some discrepancy between synthetic magnitudes and measured magnitudes may occur due to incomplete knowledge of the filter+CCD+telescope+atmosphere transmission functions that characterize the observer's system.  For a large photometric survey we assume that the photometric system will be well characterized via calibration observations, and that this will be a negligible source of error - SN colors can be calculated for an arbitrary filter system.

\begin{figure}
\epsscale{1.0}
\plottwo{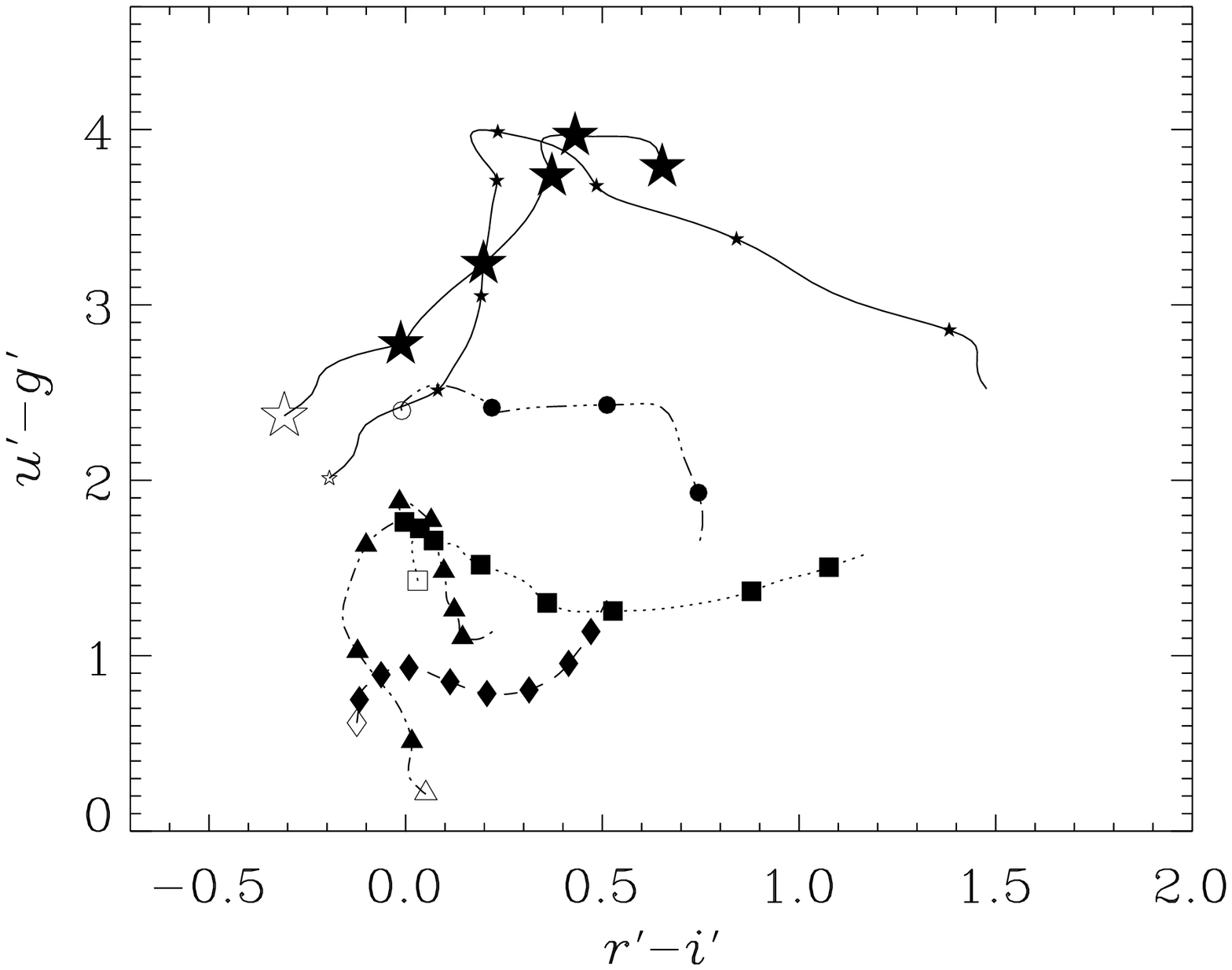}{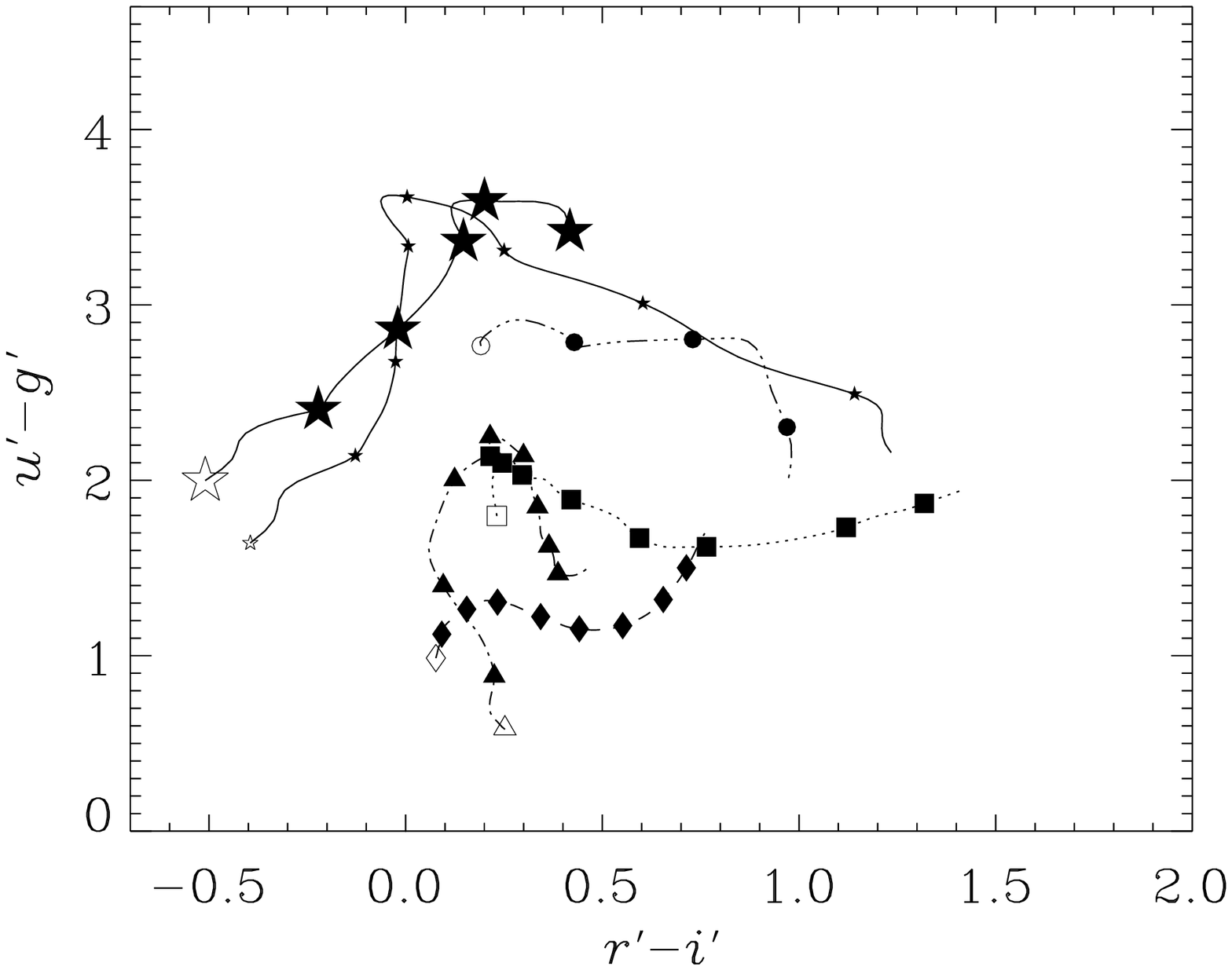}
\figcaption{{\it Left}: The $u'-g'$ vs. $r'-i'$ SDSS color-color diagram with A$_{V}$=1 mag reddening (\citet{calzetti94} dust law) applied only to the restframe Type Ia SN spectrum.  Symbols are as in Figure \ref{fig:sdss}. {\it Right}: The $u'-g'$ vs. $r'-i'$ color-color diagram with A$_{V}$=1 mag reddening (\citet{calzetti94} dust law) applied to the restframe spectra of all SN types \emph{except} Type Ia.\label{fig:red}}
\end{figure}

We must also consider reddening. Reddening can affect $\delta c$ in two ways.  First, if the reddening correction applied to the template SN spectra is incorrect, then the colors of the template spectrum will be different from that of the `true' unreddened template for a SN type.  We refer the reader to the references listed in Tables \ref{table:uvo} and \ref{table:speclist} for a discussion of the SN template reddening. Second, observed SNe will be affected by various amounts of reddening, and will thus have colors which are redder than the 'true' unreddened template SN. We have investigated the effect of reddening in the SN host galaxy by applying A$_{V}$=1 mag of dust to the \emph{restframe} template spectra using both a \citet{CCM89} dust law (containing the $2175\AA$ bump) and a \citet{calzetti94} starburst galaxy dust law. This is repeated for all redshifts studied here. These reddening vectors are shown in the color-color diagrams as arrows.  Strictly speaking, the change in color for a given amount of extinction depends on the spectrum of the object, but we find that one reddening value for each color and redshift describes the effect of reddening for all the template spectra to within 0.05 mag.  To more clearly show the effect of reddening on Type Ia SN  separation we have also created color-color diagrams with A$_{V}=1$ reddening applied to just the Type Ia SN template (Figure \ref{fig:red}$a$) and with the reddening applied to all the SN types \emph{except} Type Ia (Figure \ref{fig:red}$b$).

Lastly, for all SN types, contamination by host galaxy light may play a role.  This will depend strongly on the color and magnitude of the local background due to the host galaxy. An investigation of the demographics of host galaxies and the distribution of location of SNe within the hosts is beyond the scope of the present study.

There are two more important effects which can add to $\delta c$.  The first is the intrinsic variation in the spectra of SNe of a given type at a given epoch, referred to hereafter as spectral inhomogeneity.  The second effect is due to differences between the epoch of the template spectrum and the epoch of the observed SN, since the spectra evolve with time, and is referred to below as spectral evolution.  These effects are type dependent.  Morever, they are very difficult to quantify with the small number SN spectra presently available, especially for the core-collapse SNe.

\subsubsection{Type Ia}

We first discuss the expected spectral inhomogeneity of Type Ia SNe  based on published restframe Johnson-Cousins photometry.  \citet{nobili03} have examined a large sample of well-observed Type Ia SNe in detail.  Assuming that SN Ia colors are nearly identical at $\tau \sim +35$ days \citep{lira95} to correct for extinction, and correcting the time axis for the stretch parameter $s$ \citep{perlmutter97} they determine intrinsic dispersions in B$-$V, V$-$R, and R$-$I color.  From \bmaxr ~ to $\tau = 15$ the dispersions in all three colors are $\le 0.11$ mag. Thus we see that intrinsic differences in the restframe optical color at a single stretch corrected epoch are small, and it seems likely from the work of \citet{phillips99} and \citet{cmagic} that much of the remaining difference is correlated with the absolute magnitude of the SN (which is related to $s$). To be conservative we use twice the dispersion given by \citet{nobili03} as the contribution to $\delta c$ for all colors (adjacent bands) and redshift combinations that sample the restframe 4000-9000\AA region of the SN spectrum. 

To estimate the effect of Type Ia spectral evolution using published restframe Johnson-Cousins photometry we examine the change in color as a function of time. From \bmaxr$+5$ to \bmaxr$+20$ the rate of change of the B$-$V, V$-$R, and R$-$I colors is small, $\frac{\Delta (Color)}{\Delta t} \le 0.05$ mag/day.  \citep{nobili03}.  Differences between the true restframe epoch of a SN observed at \rmax+8 and the epoch of the template spectrum can arise for several reasons.  First, there is a randomly distributed uncertainty in the epoch of the observed SN due to the difficulty of determining the date of maximum light from a lightcurve with only a simple analysis. This uncertainty depends strongly on the specific observing scheme used and realistically will also have some dependence on the redshift of the SN, but for most surveys should be $\lsim 2$ days.  There are also errors in the epoch associated with the redshift effects described at the beginning of \S~\ref{sec:color}, which for Type Ia SNe will certainly be $\le 2$  restframe days.  The uncertainty in the epoch of the template spectrum is $\pm 2$ days.  Adding the known errors in quadrature we obtain a typical difference between the template epoch and the observed SN epoch of $\pm 4$ restframe days, corresponding to an uncertainty in color of $\pm 0.16$ mag.  We note that for Type Ia SNe an additional systematic source of uncertainty in the epoch is the lightcurve stretch parameter $s$ (or $\Delta m_{15}(\mbox{B})$), which for our purposes can be thought of as changing the spectral epoch.  This was accounted for in the analysis of \citet{nobili03}, so we must reintroduce the effect of this parameter on color assuming that $s$ will not be known from simple, preliminary lightcurve analysis (especially if the redshift is unknown!).  The template SN Ia, SN1992A, had a stretch parameter of 0.84.  Typical values of the stretch parameter range from 0.70 to 1.15, corresponding to a difference from the template epoch of $\le 2.5$ days at $\tau=9$.  This would give an additional 0.1 mag difference in color (at least for the restframe optical colors). 

\begin{figure}
\plottwo{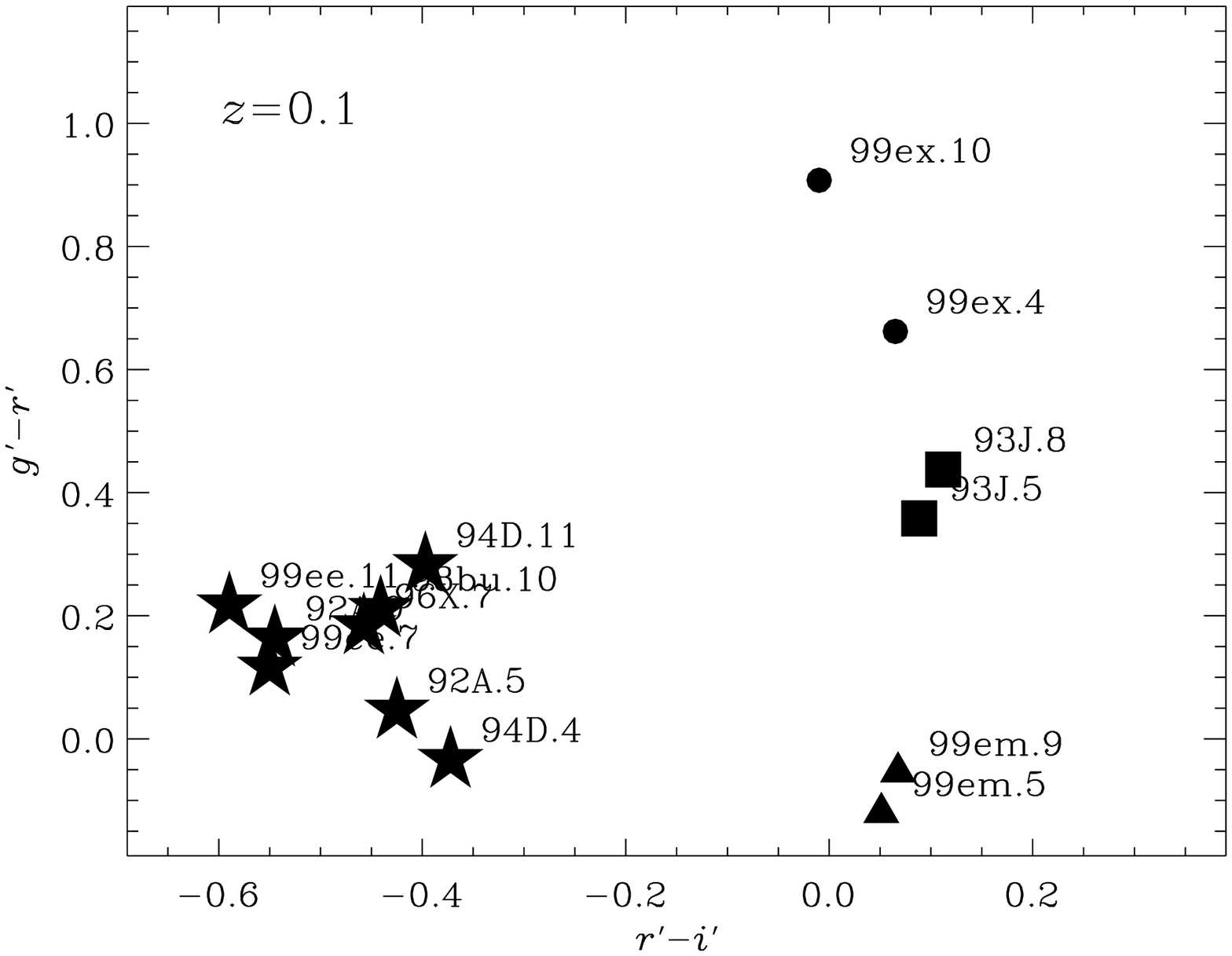}{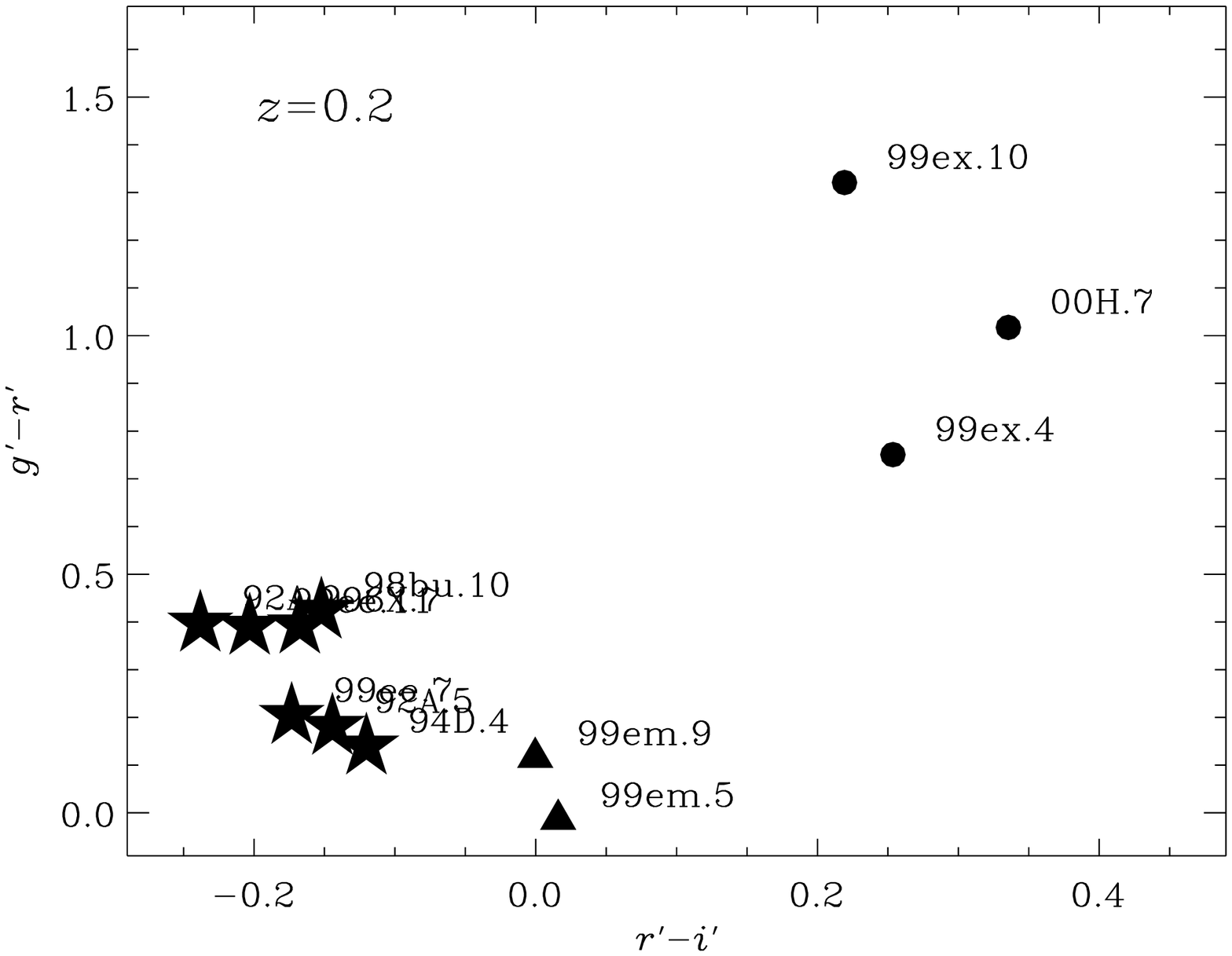}
\figcaption{{\it Left}: The $g'-r'$ vs. $r'-i'$ color-color digaram at $z=0.1$ for SNe after maximum. The symbols encode the different SN types: star~-~Ia, circle~-~Ib/c, triangle~-~IIP, square~-~IIb or IIL. The number next to each SN indicates the epoch $\tau$. {\it Left}: The same but at $z=0.2$. \label{fig:grri2max}}
\end{figure}

However, to investigate spectral inhomogenity and evolution in more detail and for the filter systems under consideration we calculate the colors of a number of different Type Ia SN spectra at epochs close to the template UV+O spectrum epoch. As an example Figure \ref{fig:grri2max} shows the $z=0.1$ and $z=0.2$ color-color diagram for 5 different Type Ia SNe (and other SN types) at $4\le \tau \le 13$.  The range in color of these spectra should therefore be caused by spectral evolution and spectral inhomogeneity, and to some extent errors in reddening determinations. Note that SN 1992A at $\tau=7$ would appear to fall near the middle of the range of Type Ia colors. To be conservative, we take one half of the full range of the resulting Type Ia colors to obtain an estimate of $\delta c$. We do this for all redshifts and colors where at least 5 spectra are available.  The results are presented in Table \ref{table:data_disp}, where the number in parenthesis gives the number of spectra from different epochs and/or SNe available at that redshift.  After adding a photometric error of 0.14 mag we can compare these values with with our expectations from the restframe Johnson-Cousins photometry given above. For example, at $z=0.6$ the $r-i$ color approximates restframe B$-$V.  We see that $\delta c$ as given by synthetic photometry using several different template spectra is generally smaller than the estimate based on Johnson-Cousins photometry.  This may reflect the small number of template spectra used to determine $\delta c$, but it gives us confidence that our estimate of $\delta c$ from Johnson-Cousins photometry is not drastically low, even when applied to different redshifts and filter combinations which only approximate restframe B$-$V.  Obviously a larger number of template spectra would improve the determination of $\delta c$ from synthetic photometry.

We estimate total values for $\delta c$ (for the SDSS system) in Table \ref{table:disps}.  These estimates are based on the behaviour of the restframe Johnson-Cousins colors of Type Ia SNe, as described above. Little weight is given to the values of $\delta c$ determined directly from template spectra, as these are likely to be too low due to our small sample. It is clear that not all the errors in color are uncorrelated, or even normally distributed, but a more rigourous analysis of $\delta c$ is unwarranted given the necessary approximations.  At higher redshift and for bluer filters we do not posses sufficient photometric or spectral data in the restframe UV to estimate $\delta c$ data. However, we expect that $\delta c$ will be larger, due to both spectral inhomogeneity and spectral evolution.  This is because we are sampling the restframe ultraviolet, $\lambda<3600$ \AA, where there are many more spectral lines that can vary in strength and location depending on the metallicity and detailed evolution of the SN, causing variations in the broadband colors \citep{kirshner92a} due to spectral inhomogeneity. Note also that evolution is generally more rapid in the bluer bands, so errors in epoch will lead to larger errors in color for the bluer bands. We therefore adopt values of $\delta c$ that increase with decreasing wavelength.

\subsubsection{Type Ib/c}
Detailed color studies such as that presented in \citet{nobili03} are not available for Type Ib/c SNe.  In fact, color curves have been published for only a handful of SN Ib/c.  The three most complete published Type Ib/c color curves are collected in \citet{max}. Despite the dearth of observational data, it is often stated that SNe Ib/c fall into two groups, fast evolving (in both luminosity and color) and slowly evolving.  It is not clear whether there is a continous distribution between these extremes, or what is the range in `speed' of Type Ib/c SNe. However, as discussed in \S~\ref{sec:specIc} the difference in B$-$V at \bmaxr for the fastest and slowest known evolving Type Ib/c is approximately 0.3 mag (the range decreases slightly at $\tau =10$).  Near \bmaxr ~the rate of change of B$-$V is largest for the fast evolving Type Ib/c: $\frac{\Delta (B-V)}{\Delta t} \approx 0.1$ mag day$^{-1}$.  Near \bmaxr$+10$ the rate is less for both fast and slowly evolving Type Ib/c, approximately 0.06 mag day$^{-1}$.  This would give a $\pm 0.3$ magnitude error in color for $\pm 5$ day error in epoch.

We can use these data to estimate $\delta c$, as was done above for Type Ia SNe.  The contribution of spectral inhomogeneity must be carefully considered, since our template spectrum is a hybrid spectrum formed from a fast evolving SN in the ultraviolet and a more slowly evolving SN in the optical.  As already discussed, this will cause the colors calculated from these different regions of the spectrum to be inconsistent with each other.  However, an adequate estimation of $\delta c$ will insure that the correct, consistent SN colors will fall within the quoted range of color.  We thus use the \emph{full} range in B$-$V color for well-observed Type Ib/c SNe as the contribution of spectral inhomogeneity to $\delta c$ for the appropriate redshifts and filters. Adding to this the contribution of spectral evolution noted above and photometric errors then gives an estimate for $\delta c $ of approximately 0.5 mag.  This will be an appropriate value for $\delta c$ when the redshift an filter combination samples the restframe B and V region of the spectrum, i.e. $\delta(b-r)$ for $z \le 0.3$ and $\delta (r-i)$ for  $0.3 < z \le 0.6$.  For colors and redshifts which sample bluer regions of the spectrum we again adopt values of $\delta c$ that increase with decreasing wavelength.

\subsubsection{Type II}
\label{sec:dispII}
We would need many UV+O spectra spanning a range of epochs for many Type II SNe of different subclasses to adequately investigate the contribution of spectral inhomogeneity and spectral evolution to $\delta c$.  As these data do not exist we must rely on studies of the optical photometric behaviour of Type II SNe, the most comprehensive of which is presented in \citet{patat}.  We can simultaneously consider the effects of spectral inhomogeneity and spectral evolution by examining their Figure 3, where they present B$-$V color curves of 21 Type II SNe including Type IIP, Type IIL, and Type IIpeculiar (no correction for reddening has been applied).  Spectral inhomogeneity is accounted for by including 18 SN color curves (we ignore SNe  1972Q, 1973R, and 1987A which are all anomalously red; two of these are likely highly reddened and we show the effects of reddening elsewhere) and determining the width of the B$-$V color curve.  We can simultaneosly take into account the effect of spectral evolution by then considering the full range of colors for these 18 SNe from $\tau =0$ to $\tau=30$ days. We use a large range of $\tau$ to account for the difficulty in determining epochs for Type II lightcurves, especially Type IIP. This gives a $\pm 0.5$ mag range in color about the central value. The range in epoch that we have used for Type IIP is a pessimistic estimate when applied to Type IIb and IIn, but we use the these ranges as a safe upper limit to $\delta c$.  This is especially true of the Type IIb template colors, which were calculated using a $\tau=0$ spectrum. Note also that the colors of the Type II template spectra that we use are somewhat bluer than the central value: this is to be expected when using a range in $\tau$ that extends to such late epochs while the template spectra come from very early epochs.

\subsection{Discussion of Color Based Separation}
\label{sec:discus}
\begin{figure}
\plotone{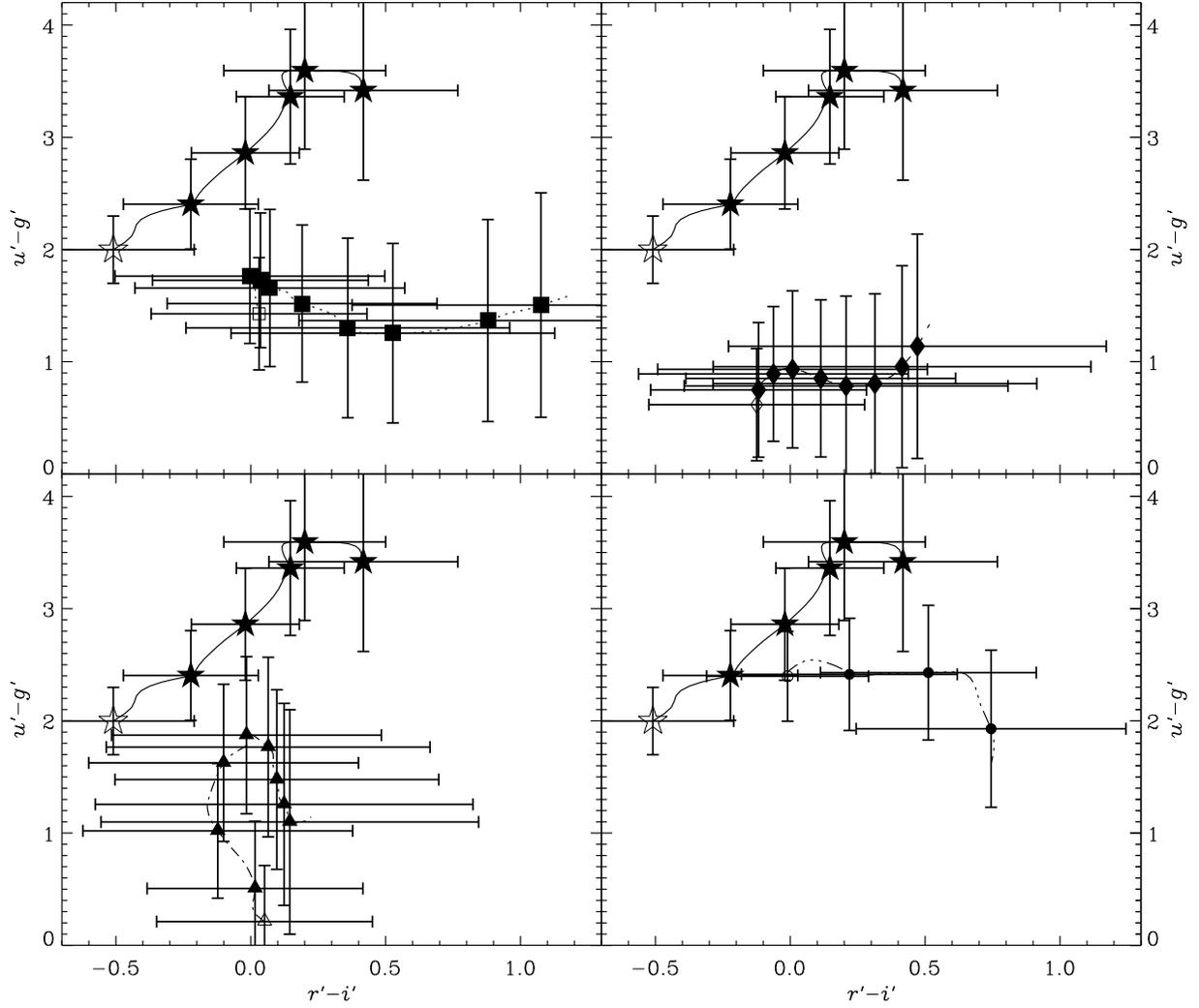}
\figcaption{Type Ia SN ($\tau=10$) synthetic colors ($u'-g'$ vs. $r'-i'$) compared to other types, with error bars indicating the expected scatter in observed SN colors (from Table \ref{table:disps}).  The symbols are as in Figure 3. {\it Top Left}: Type Ia colors vs. Type IIb colors. {\it Top Right}: Type Ia colors compared to Type IIn colors. {\it Bottom Left}: Type Ia colors compared to Type IIP colors. {\it Bottom Right}: Type Ia colors compared to Type Ib/c colors. \label{fig:ugri_multi}}
\end{figure}

In the ideal case where $u$, $b$, $r$, and $i$ (or $u'$, $g'$, $r'$, and $i'$) magnitudes are available we see that the separation of SN Type Ia from other types is very good, even when we take into account realistic errors in the determination of the $r_{max}+8$ day colors (Figure \ref{fig:ugri_multi}). The most likely contaminant is low redshift Type Ib/c SNe, which may be confused with higher redshift Type Ia given the expected errors in color.  Other possible contaminants include high-redshift ($z > 0.4$) Type IIP SNe and moderate redshift Type IIb SNe.  In Figure \ref{fig:red} we see that due to the direction of the reddening vectors, reddening does not appear to pose a significant problem for Type Ia SN identification in the $u-b$ vs. $r-i$ diagram.  The only exception is that reddened low redshift Type Ia will look like higher redshift Type Ia, and can thus possibly be confused with low redshift Type Ib/c as already mentioned.  This will not affect the absolute rate of contamination of the Type Ia sample by Type Ib/c.  To identify SNe which are likely to be of Type Ia using this diagram one can choose SNe which are very red in $u-b$ color for a given value of $r-i$.

When fewer colors are available the difficulty of Type Ia SN identification significantly increases.  In the $u-b$ vs. $b-r$ diagram it is clear that while the contaminants will be similar the amount of overlap between Type Ia SNe and these other types will increase, making it more likely that SNe are misidentified. It also appears that reddening begins to play more of a role.  This is especially true of the $b-r$ vs. $r-i$ diagram, where the only unambiguous Type Ia SNe are at $z \le 0.2$, and these SNe can easily be reddened into regions of color-color space occupied by other SN types.  However, if a SN is observed to have $r-i \lsim -0.1$ mag then it is likely to be a Type Ia SN at very low redshift.

\section{Adding Magnitude Information: Color-Magnitude Diagrams}
\label{sec:mag}
It appears possible to use $u-b$ vs. $r-i$ diagrams to aid in the selection of probable Type Ia SNe, since many regions in color-color space that are occupied by non-Ia are not occupied by Type Ia at any of the redshifts considered here.  Thus, objects populating these regions can be excluded from the SN Ia sample without any further information.  But can we do better? Type Ia SNe have reasonably constant luminosity, therefore it might be useful to look at magnitude information as an additional discriminator.  We here consider the addition of magnitude information in the form of color-magnitude diagrams. 

We created color-magnitude diagrams of the template spectra in a similar manner to the way we created color-color digrams in \S~\ref{sec:color}. The flux scale of the spectra was calibrated so that the SNe have typical luminosities for their type \citep{Richardson02} and age.

\subsection{Errors in the Diagrams}
We have already examined the expected typical spread in color for the SNe types, $\delta c$.  Here we examine the dispersion in magnitude expected at a given redshift. Extinction will have the same effect in the magnitude axis that reddening does on the color axes.  Additionally, amplification by gravitational lensing can affect the observed magnitude, but this is expected to be a neglible effect at the redshifts investigated here.  By far the largest effect, for all SN types, is the intrinsic dispersion in SN luminosity.  \citet{Richardson02} have examined this phenomenon for many SN types and we use their results. Their central absolute B magnitude values for each type, rescaled to H$_{\circ}=70$ km s$^{-1}$ Mpc$^{-1}$, converted to AB magnitude, and combined with template lightcurves (or the published $m-m_{max}$ in the case of Types IIn and IIb), were used to flux calibrate the template UV+O spectra from Table \ref{table:uvo}.  In the case of the Type IIb template (not considered by \citet{Richardson02}) we have used a luminosity appropriate for a normal Type Ib/c or a normal Type IIL.  We do not add magnitude differences caused by uncertainty in the lightcurve epoch or measurement errors, and we use the $1\sigma$ standard deviations of \citet{Richardson02}.  The magnitude errors presented should thus be taken as \emph{lower} limits.

\subsection{Discussion of Color-Magnitude Based Separation}
\begin{figure}
\plotone{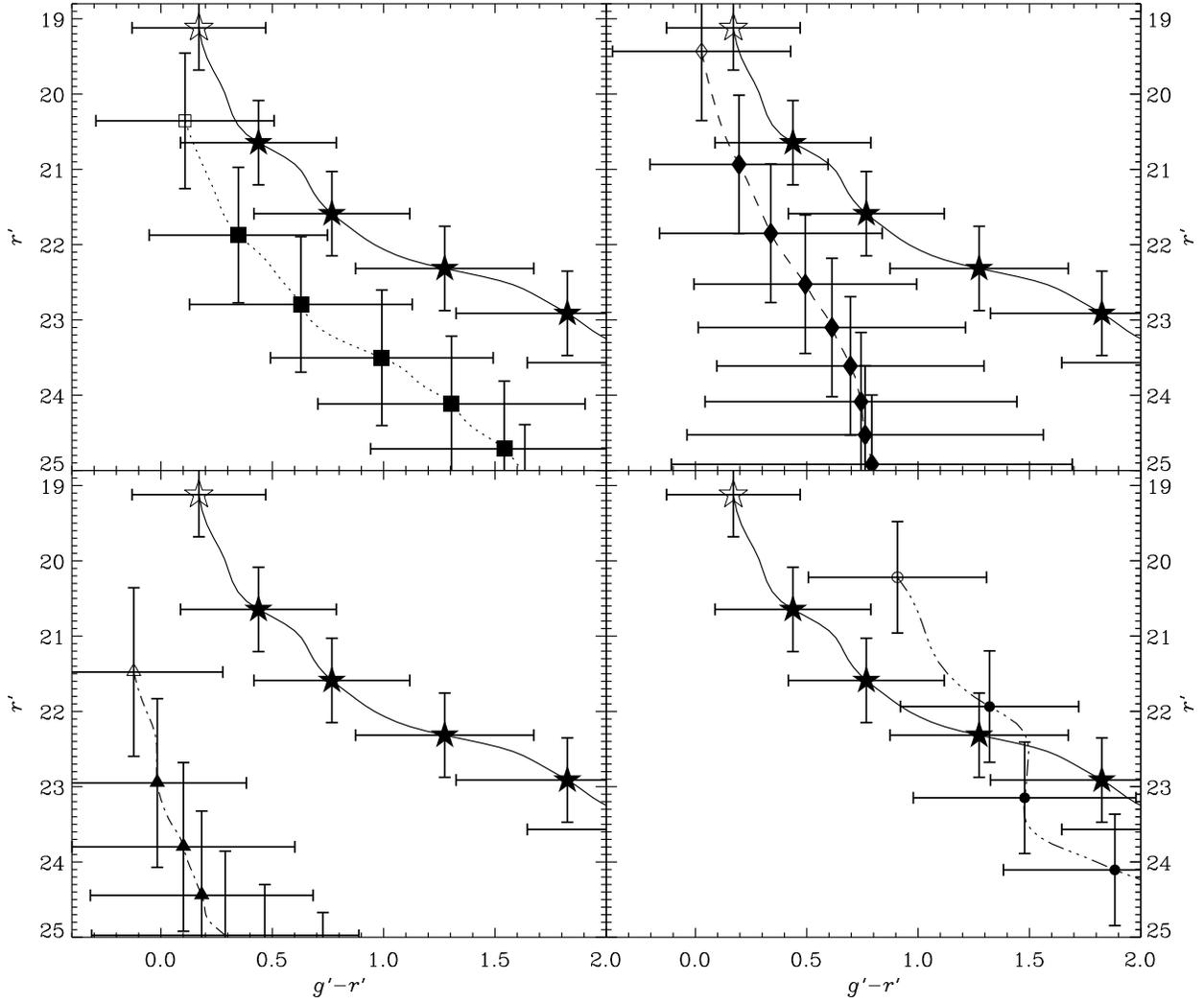}
\figcaption{The color-magnitude diagram of the template SNe for the SDSS filters, with magnitude dispersions from \citet{Richardson02}.   Symbols are as in Figure \ref{fig:sdss} \label{fig:cmd}}
\end{figure}

Figure \ref{fig:cmd} shows that there is an enormous amount of overlap between SN Ia and other SN types in the color-magnitude diagram.  Especially problematic is the continued confusion of low redshift SNe Ib/c with higher redshift SNe Type Ia.  The essential problem is that while Type Ib/c SNe are redder than Type Ia at a given redshift, they are also intrinsically fainter.  SNe Ib/c at $z=0.1-0.3$ have similar colors as higher redshift Type Ia and are intrinsically less luminous by an amount that is similar to the distance modulus from $z$=0.2 to $z$=0.4.  Reddening and extinction by dust of $z=0.1-0.2$ Type Ia's can make them look like these low redshift Type Ib/c, and reddening and extinction of these low redshift Type Ib/c will make them look like even higher redshift ($z \geq 0.5$) Type Ia.  Without independent redshift information there is then a problem in separating Type Ia from some Type Ib/c SNe using only color and magnitude information at a single observed epoch.  This is a generic problem that will occur for any combination of filters and SN redshift which samples the same wavelngth region in the restframe spectrum.
  
However, Figure \ref{fig:cmd} does show that high redshift Type IIP, which have similar colors to low redshift Type Ia, are too faint to be confused with these Type Ia, if indeed they are even detected.  Even if we account more rigorously for dispersion in magnitude, using twice the standard deviation of \citet{Richardson02}, adding a typical photometric error of 0.1 mag., and adding a magnitude error due to uncertainty in the lightcurve epoch we find that magnitude combined with color information at a single known epoch can be used to separate Type IIP SNe from Type Ia. Even if Type Ia SNe are extincted by dust so that they have the same magnitude as these high redshift Type IIP, they will be reddened so that they have different colors. Thus one of the contaminants given in \S~\ref{sec:discus} has been removed.

\section{Adding Temporal Information}
\label{sec:time}
We have examined SN colors only at one epoch, $\tau \sim 10$, because of the availability of UV spectra at that epoch (with the notable exception of Type IIb and Type IIL SNe).  However, it is possible that similar color based identification of Type Ia SNe can be accomplished by looking at colors at other epochs, for example at maximum light or even before maximum.  This would have the advantages that spectroscopic followup could be scheduled at or very close to maximum light, and that the SN would be brighter and detectable to a larger redshift.  Use of the time observable in conjunction with color or magnitude information may also provide another way to separate SN types if their evolution in color or magnitude space is unique.  This may be especially useful when using $b-r$ vs. $r-i$ colors, where identification of Type Ia based solely on colors at one epoch is difficult.To investigate these questions we examine the time evolution of SN colors. Much of the following material has been presented in \citet{poznanski}.  We here change only the filters, and restrict the SN epoch to less than 35 days after \bmaxr.

\subsection{Color-Color as a Function of Time}
Because of the significant confusion in the $b-r$ vs. $r-i$ diagram it is worthwhile to look deeper for methods to disentangle these populations using only photometry.  While color-magnitude diagrams help with the problem of Type IIP SNe, we also consider the evolution of color with epoch, i.e. the paths traced out in the color-color diagram as a function of epoch, as a way to discriminate between SN types.

\begin{figure}
\plottwo{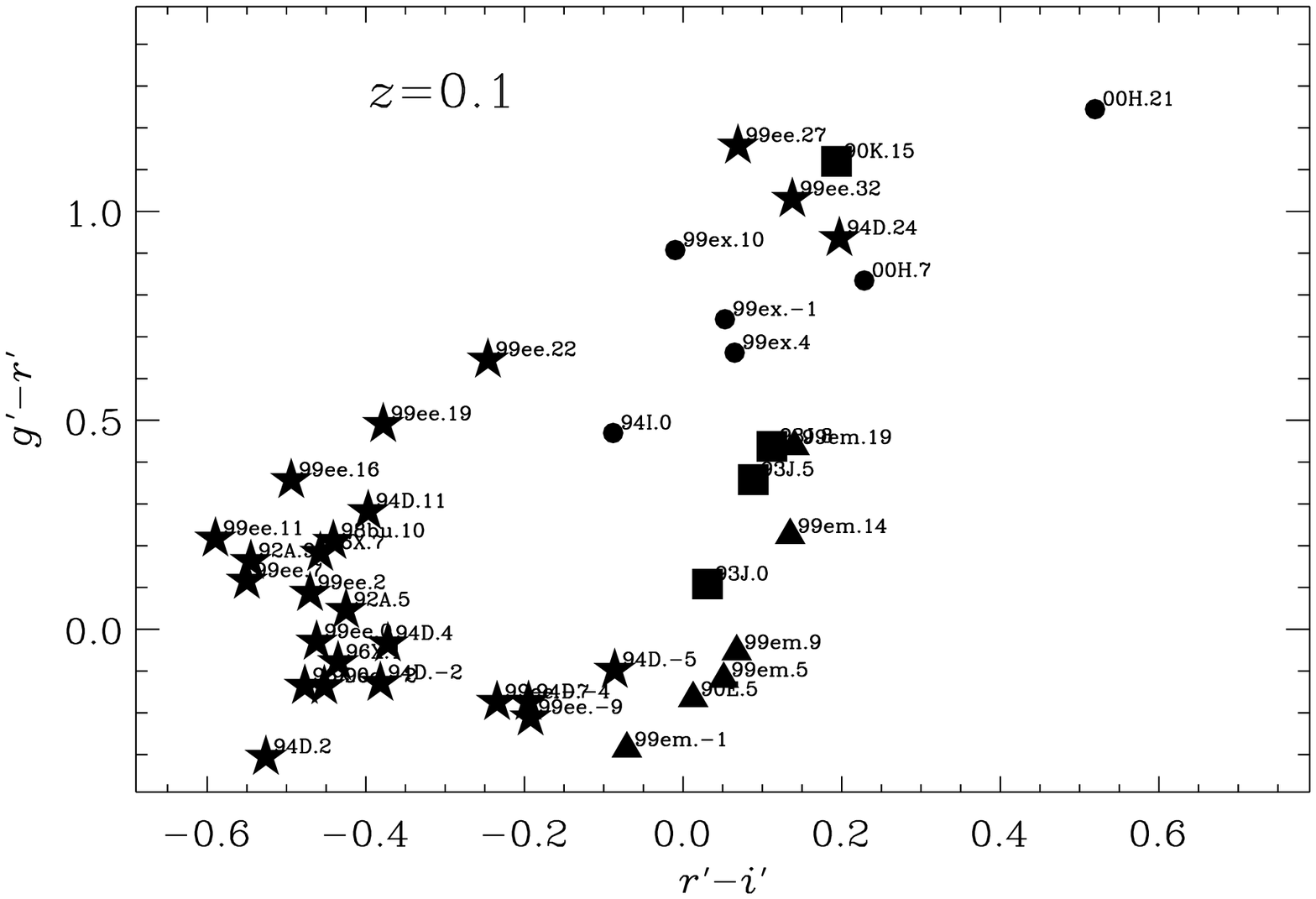}{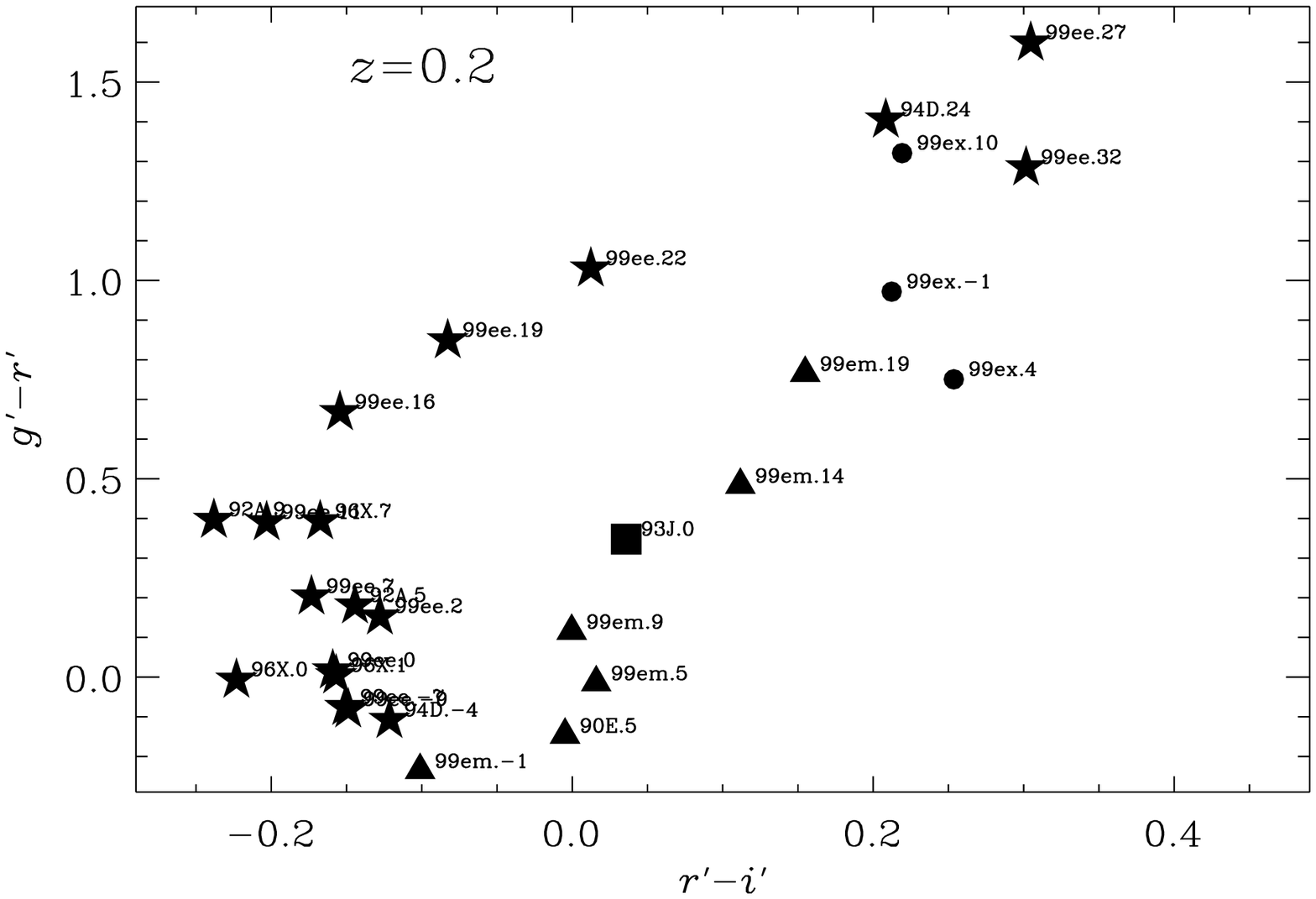}
\figcaption{{\it Left}: The $b-r$ vs. $r-i$ color-color digaram at $z=0.1$ for all SN templates with $\tau < 36$.  The number next to each type indicates the epoch. {\it Right}: The same but at $z=0.2$.\label{fig:grri_all}}
\end{figure}

We have sufficient spectral and epoch coverage for a large number of SNe to investigate their paths in the $b-r$ vs. $r-i$ color-color diagram at redshifts $z=0.1$ and $z=0.2$ (Figure \ref{fig:grri_all}).  This is not true of higher redshifts, where there is insufficient restframe UV spectral data.  This example must then be taken only as indicative of the further information about SN type which may be gleaned from the full multi-color light curve information. As more data become available we will be able to extend the use of color-color-time information to other filters and redshifts.  We see, however, that at these low redshifts SNe Ia follow a particular path in color-color space, as shown by \citet{poznanski} for Johnson-Cousins filters.  They become bluer in $r-i$ as they rise toward maximum, become redder in $b-r$ as they pass maximum, and start to become redder in $b-r$ and $r-i$ about 15 days after \bmaxr. There is a hint that Type Ib/c and Type IIb have a different evolution, moving to redder colors in both $b-r$ and $u-b$ as they evolve from just before maximum to after maximum.  Similar trends are seen in the $z=0.1$ $u-b$ vs. $b-r$ and $b-r$ vs. $r-i$ diagrams. It may thus be possible, as \citet{poznanski} suggest, to use the shape of the color evolution tracks to help discriminate between SN types, though this will require information obtained at a large range of epochs. We have not been able to investigate how the shape of these tracks changes at higher redshifts, which could complicate the method without independent redshift estimates. It appears from Figure \ref{fig:grri_all} that type separation at early epochs will be more difficult, since the early colors of Type Ia are closer to the early colors of other SN types at this redshift, not to mention the early colors of SNe at different redshifts.  However, confirmation of this will have to await more data on the ultraviolet evolution of SN, so that the analysis can be carried out at more redshifts and for more SNe.

\subsection{Magnitude as a Function of Time, i.e. the Lightcurve}
Throughout the above discussion we have implictly assumed that some information about the lightcurve is available, specifically the date of maximum light through a certain filter.  If this is not true then other information will be necessary to identify Type Ia, such as an estimate of the redshift as suggested in \citet{poznanski}.  But to what extent does the lightcurve itself constrain the SN type?  We have already said that the light curves of Type IIP events are significantly different in shape from Type Ia lightcurves, marked by a rapid rise time and, by definition, a relatively slow decay \citep{leibundgut91}, especially in the redder bands.  Indeed, this has been utilized informally by \citet{barris_ifaSn}. However, in the restframe U, B and V bands the decay is much more rapid and the lightcurve of moderately redshifted events may superficially resemble those of SNe Ia.  It is difficult to distinguish between any of the other types based on a simple lightcurve analysis near maximum. While the different types have different decline rates \citep{richmond96}, type separation based on this feature is complicated by the effect of cosmological time dilation.

\section{On the Use of Host Galaxy Characteristics}
\label{sec:host}
We have assumed above that the redshift of the SN cannot be determined independently.  However, photometric redshifts of the host galaxies may provide a redshift constraint that would be useful in SN type identification. Additionally, if the Hubble type of the host can be determined, either morphologically or jointly with the photometric redshift estimate from galaxy spectral energy distribution fitting, then we will have a significant clue to the SN type.  This is due to the extreme scarcity of Type II and Type Ib/c SNe in E/S0 galaxies.  The two bluest filters in the SDSS and logarithmic systems are very sensitive to the redshifting of the 4000 \AA  ~break, which is important to the determination of the galaxy photometric redshift and galaxy type.

While the $(1+z)^{4}$ cosmological surface brightness dimming is beneficial in reducing the photon background from the SN host galaxy, it makes detecting the host galaxy difficult. At faint magnitudes assigning a SN to a unique host galaxy may become problematic as well.  It will be important to examine the redshift limit to which photometric redshifts and morphologies can be accurately determined.  This limit will certainly be lower than the detection limit for Type Ia SNe unless coadded images from many different nights are used for photometry of the host galaxies.  While photometric redshifts are typically not accurate to better than $\Delta z=0.1$ \citep{hogg} even this accuracy is sufficient to distinguish low redshift Type Ib/c from higher redshift Type Ia SNe.  Also, removing this degeneracy requires photometric redshifts of host galaxies to only $z=0.3$.  To the extent that independent redshift information for the hosts is available the problem of type separation is reduced to that presented in \citet{poznanski}, and it is possible to relax requirements on the knowledge of, e.g. the epoch of the SN.  We note, however, that the use of blue magnitudes at moderate redshift ($z \geq 0.3$) significantly increases the ease of type separation.

\section{Conclusions}
We have shown that by using photometric information over the entire optical spectrum it is possible to separate Type Ia SNe from most other SN types simultaneously over a large range in redshift. This can be done using magnitude and color information from only a single observed epoch close to maximum light (\rmax +8), provided that the epoch is known to reasonable accuracy.  There is still some confusion with other SN types, particularly Type Ib/c, and we suggest that this can be reduced by examining additional epochs and noting the color evolution of the confusing events, and by obtaining photometric redshifts of the host galaxies.  It is also possible that type separation can be accomplished at epochs other than those studied here.  However, the use of very blue magnitudes appears to be crucial for separation of Type Ia SNe from other types at $z > 0.2$, at least without extensive temporal information or a constraint on the SN redshift. This is troubling since the very characteristic of Type Ia SNe which makes blue magnitudes useful, their faintness in this wavelength region, also makes photometry difficult.

It is important to emphasize that we have attempted to estimate realistic values of the dispersion in SN colors, but that we are greatly hindered by the lack of restframe ultraviolet spectra.  While we have noted the relevant sources of $\delta c$ in \S~\ref{sec:disp}, in many cases the available data do not allow us to precisely determine the error introduced by each source.  The errors that we have given in many cases reflect a lack of knowledge rather than an irreducible scatter in measurements.  Also, many sources of error depend upon the specific observing strategy that is used. We believe that the color dispersions are small enough to allow type separation in many cases, but more ultraviolet spectroscopy (or even photometry!) of different SNe at a variety of epochs is necessary to confirm this claim. Further extensions of this method are the inclusion of the spectra of peculiar Type Ia SNe and hypernovae, and spectra \emph{at} SN maximum light.  Taken to its logical conclusion the method presented here could be used with a spectral database covering many SNe of different subtypes over a large, well-sampled range of epochs and ultraviolet wavelengths to construct synthetic optical magnitudes as a function of redshift, time, color, and SN type for any filter system.  Such a multidimensional parameter space could then be collapsed and/or binned in any dimension or dimensions to reflect the realities of a given survey or observation, and the resulting parameter space searched for the best photometric indicators of SN type while accurately characterizing the typical dispersion in colors.

The criteria presented here (and in \citet{poznanski}) for the photometric identification of Type Ia SNe would allow the efficient targeting of Type Ia SNe discovered by imaging surveys for spectroscopic and photometric followup.  A sample of Type Ia SNe derived from such surveys will have many uses.  First, the method outlined here could be tested and improved, and errors quantified more precisely. This would allow yet more efficient spectroscopic followup in future large imaging surveys such as those conducted with LSST. There is the potential for a large increase in our knowledge of the photometric phenomenology of all SN types, especially regarding their properties in the ultraviolet.   Second, other new parameters related to Type Ia luminosity may be identified.  For example, the large synoptic surveys will be very sensitive to the rise time of a SN, at least for low redshift Type Ia SNe. This is thought to be related to the intrinsic luminosity of SNe Ia \citep{reiss_rise}.  The ultraviolet behaviour of SNe may be important for investigating the effects of metallicity or dust on the luminosity of Type Ia SNe. Third, if one accepts the systematic effects as negligible, then it is possible to use a large sample of Type Ia SNe to confirm and refine measurements of cosmological parameters, especially $\Omega_{\Lambda}$ and $w$, the equation of state parameter of the dark energy.  Some authors have suggested that time-consuming spectroscopic followup to obtain redshifts may not be necessary for such a study \citep{barris}, but reliable photometric identification of the Type Ia SNe, using methods such as those outlined here, will be crucial.

\acknowledgements
First, we wish to acknowledge the many people who have spectroscopically observed SNe and made their observations publicly available, especially Mario Hamuy and the Asiago SN group.  We also wish to thank the SINS group for carrying out ultraviolet spectroscopy of different SN types over many years, using IUE and HST. Finally, we would like to thank Kim Shella for useful discussions.  B.D.J. was supported by NSF grant AST 02-06048, and we acknowledge support by NASA grant NAG5-13081 and STScI grant GO 10182.  This work has made use of the exceptional SUSPECT Database, compiled and maintained by Dean Richardson at the University of Oklahoma.  Some of the data presented in this paper were obtained from the Multimission Archive at the Space Telescope Science Institute (MAST). STScI is operated by the Association of Universities for Research in Astronomy, Inc., under NASA contract NAS5-26555.

\clearpage
\bibliographystyle{apj}		
\bibliography{apj-jour,Johnson}


\clearpage

\clearpage
\begin{deluxetable}{cccccccc} 
\rotate
\tablecolumns{8} 
\tablewidth{0pc} 
\tablecaption{UV+O SN Templates - Good UV Coverage \label{table:uvo}} 
\tablehead{ 
\colhead{SN Name} & \colhead{Type}   & \colhead{Redshift}& \colhead{UT date$_{Bmax}$}    & \colhead{epoch ($\tau$)} & \colhead{$\lambda\lambda$}    & \colhead{A$_{V}$}   & \colhead{Ref.}   }
\startdata 
1992A       & Ia & 0.0062     & 1992 Jan 19       & 5, 9& 2000-9820 &0.1 &1\\
1993J       & IIb& -0.0001    & 1993 Apr 15       & 0    & 1600-9910 &1.0 &2,3,4,5\\
1994I+1999ex& Ic &0.0015,0.011& 1994 Apr 8        & 10   & 2250-9650 &1.2,1.0\tablenotemark{a}&6,7,8\\
1998S+b-body& IIn& 0.0028     & $\sim$1998 Mar 20 & 10   &1200-10000 &0.6 &9,10,11,12\\
1999em      & IIP& 0.0024     & 1999 Oct 31 & 5\tablenotemark{b}&1200-10050&0.3& 13,14,15,16\\
\enddata
\tablenotetext{a} { The two values refer to SN1994I and SN 1999ex respectively.}
\tablenotetext{b}{ The optical portion of the spectrum redward of 7700 \AA ~was formed by interpolating between the $\tau=+3$ and $\tau=+8$ spectra of reference (14).}
\tablerefs{(1) \citet{kirshner92a}; (2) \citet{filippenko93J}; (3) \citet{jeffery93J}; (4) \citet{asiago93J}; (5) \citet{93Jphot}; (6) \citet{millard94I}; (7) \citet{IcSpec}; (8) \citet{94Iphot}; (9) \citet{leonard2000}; (10) \citet{fassia_spec}; (11) \citet{fassia_phot}; (12) \citet{anupama}; (13) \citet{baron2000_99em}; (14) \citet{hamuy99em}; (15) \citet{leonard_99em}; (16) \citet{elmhamdi_99em}}
\end{deluxetable}

\clearpage
\begin{deluxetable}{cccccc} 
\tablecolumns{6} 
\tablewidth{0pc} 
\tablecaption{SN Templates - Poor UV coverage\label{table:speclist}}
\tabletypesize{\scriptsize} 
\tablehead{
\colhead{SN Name}  &  \colhead{Type}  &  \colhead{Epoch ($\tau$)}  &  \colhead{Redshift}  &  \colhead{A$_{V}$} & \colhead{Ref}}
\startdata
92A & Ia &        5 &    0.0062 &      0.1 & see Table \ref{table:uvo}\\
92A & Ia &        9 &    0.0062 &      0.1 & see Table \ref{table:uvo}\\
94D & Ia &      -11 &    0.0015 &      0.1 & 2\\
94D & Ia &       -9 &    0.0015 &      0.1 & 2\\
94D & Ia &       -5 &    0.0015 &      0.1 & 2\\
94D & Ia &       -4 &    0.0015 &      0.1 & 2\\
94D & Ia &       -2 &    0.0015 &      0.1 & 2\\
94D & Ia &        2 &    0.0015 &      0.1 & 2\\
94D & Ia &        4 &    0.0015 &      0.1 & 2\\
94D & Ia &       10 &    0.0015 &      0.1 & 2\\
94D & Ia &       11 &    0.0015 &      0.1 & 2\\
94D & Ia &       24 &    0.0015 &      0.1 & 2\\
96X & Ia &       -2 &    0.0089 &      0.2 & 3\\
96X & Ia &        0 &    0.0089 &      0.2 & 3\\
96X & Ia &        1 &    0.0089 &      0.2 & 3\\
96X & Ia &        7 &    0.0089 &      0.2 & 3\\
98bu & Ia &      10 &     0.003 &      1.0 & 4,5\\
99ee & Ia &      -9 &     0.011 &      0.9 & 6,7\\
99ee & Ia &      -7 &     0.011 &      0.9 & 6,7\\
99ee & Ia &      -2 &     0.011 &      0.9 & 6,7\\
99ee & Ia &       0 &     0.011 &      0.9 & 6,7\\
99ee & Ia &       2 &     0.011 &      0.9 & 6,7\\
99ee & Ia &       7 &     0.011 &      0.9 & 6,7\\
99ee & Ia &       9 &     0.011 &      0.9 & 6,7\\
99ee & Ia &      11 &     0.011 &      0.9 & 6,7\\
99ee & Ia &      16 &     0.011 &      0.9 & 6,7\\
99ee & Ia &      19 &     0.011 &      0.9 & 6,7\\
99ee & Ia &      22 &     0.011 &      0.9 & 6,7\\
99ee & Ia &      27 &     0.011 &      0.9 & 6,7\\
99ee & Ia &      32 &     0.011 &      0.9 & 6,7\\
\tableline
94I & Ic &        0 &    0.0015 &      1.2 & 8\\
99ex & Ic &      -1 &     0.011 &      1.0 & 6,7\\
99ex & Ic &       4 &     0.011 &      1.0 & 6,7\\
99ex & Ic &      10 &     0.011 &      1.0 & 6,7\\
00H & Ib &        1 &    0.0132 &      0.0 & 9\\
00H & Ib &        7 &    0.0132 &      0.0 & 9\\
00H & Ib &       21 &    0.0132 &      0.0 & 9\\
\tableline
90E & IIP &       5 &    0.0041 &      0.5 & 10,11\\
90E & IIP &      13 &    0.0041 &      0.5 & 10,11\\
99em & IIP &     -1 &    0.0024 &      0.5 & see Table \ref{table:uvo}\\
99em & IIP &      3 &    0.0024 &      0.5 & see Table \ref{table:uvo}\\
99em & IIP &      8 &    0.0024 &      0.5 & see Table \ref{table:uvo}\\
99em & IIP &      9 &    0.0024 &      0.5 & see Table \ref{table:uvo}\\
99em & IIP &     14 &    0.0024 &      0.5 & see Table \ref{table:uvo}\\
99em & IIP &     19 &    0.0024 &      0.5 & see Table \ref{table:uvo}\\
99em & IIP &     35 &    0.0024 &      0.5 & see Table \ref{table:uvo}\\
93J & IIb &       0 &  -0.0001  &      1.0 & see Table \ref{table:uvo}\\
93J & IIb &       5 &   -0.0001 &      1.0 & see Table \ref{table:uvo}\\
93J & IIb &       8 &   -0.0001 &      1.0 & see Table \ref{table:uvo}\\
93J & IIb &     -15 &   -0.0001 &      1.0 & see Table \ref{table:uvo}\\
90K & IIL &      15 &    0.0053 &      1.5 & 12\\
\enddata
\tablerefs{(2) \citet{patat_94D}; (3) \citet{salvo_96X};  (4) \citet{cappellaro_98bu}; (5) \citet{jha_98bu}; (6) \citet{IcSpec}; (7) \citet{max}; (8) \citet{clocchiatti94I}; (9) \citet{branch_Ib2002}; (10) \citep{schmidt_90E}; (11) \citet{benetti_90E}; (12) \citet{Cappellaro_90K}}
\end{deluxetable}

\clearpage
\begin{table}
\begin{center}
\caption{Possible Sources of Difference Between Synthetic and Observed SN Colors\label{table:errors}}
\begin{tabular}{cl}
\tableline\tableline
\multicolumn{2}{l}{Photometric Errors}\\
   & \scriptsize Spectrophotometric errors in the template spectra\tablenotemark{a}\\
   & \scriptsize Errors in the adopted filter+CCD response functions\\\
   & \scriptsize Expected photometric errors of the observed SNe\\
   & \scriptsize Contamination by host galaxy light\\
\multicolumn{2}{l}{Reddening\tablenotemark{b}}\\
   & \scriptsize Incorrect dereddening of the template\\
   & \scriptsize Reddening of the observed SN\\
\multicolumn{2}{l}{Intrinsic Difference in SN Spectra}\\
   & \scriptsize Is the template spectrum typical for its type?\\
   & \scriptsize What is the intrinsic scatter about the typical spectrum?\\
\multicolumn{2}{l}{The Epoch of the SN}\\
	& \scriptsize Uncertainty in the epoch of the template spectrum\\
	& \scriptsize Error in correction for redshift effects (time dilation and change of max. light date)\\
	& \scriptsize Stretch Correction effects (for Type Ia SNe)\\
	& \scriptsize Uncertainty in the epoch of the \emph{observed} SN\\	
\tableline
\tableline
\tablenotetext{a}{Only spectrophotometric errors that are a function of wavelength are relevant.}
\tablenotetext{b}{Due to dust in the SN Host and in the Milky Way.}
\end{tabular}
\end{center}
\end{table}

%

\clearpage
\begin{table}
\begin{center}
\caption{$\delta c$ from Synthetic Photometry of Type Ia SNe\label{table:data_disp}}
\begin{tabular}{ccccc}
\tableline\tableline
 Redshift & $\delta (u'-g')$ & $\delta (g'-r')$ & $\delta (r'-i')$ \\
\tableline
   0.1 &    - & 0.15(9) & 0.10(9) \\
   0.2 &    - & 0.15(8) & 0.08(9) \\
   0.3 &    - & 0.12(3) & 0.05(9) \\
   0.4 &    - & - & 0.15(9) \\
   0.5 &    - & - & 0.22(8) \\
   0.6 &    - & - & 0.12(7) \\
\tableline
\tableline
\end{tabular}
\end{center}
\end{table}

\clearpage
\begin{table}
\begin{center}
\caption{Adopted Total $\delta c$ for All SN Types\label{table:disps}}
\begin{tabular}{ccccc}
\tableline\tableline
Type & Redshift & $\delta (u'-g')$ & $\delta (g'-r')$ & $\delta (r'-i')$ \\
\tableline   

Ia &0.1 &  0.30 & 0.30 & 0.30 \\
   &0.2 &  0.40 & 0.35 & 0.25 \\
   &0.3 &  0.50 & 0.35 & 0.20 \\
   &0.4 &  0.60 & 0.40 & 0.20 \\
   &0.5 &  0.70 & 0.50 & 0.30 \\
   &0.6 &  0.80 & 0.60 & 0.35 \\
   &0.7 &  - & 0.70 & 0.40 \\
   &0.8 &  - & 0.80 & 0.45 \\
   &0.9 &  - & 0.80 & 0.50 \\
   &1.0 &  - & - & 0.55 \\
\tableline
Ib/c & 0.1 & 0.4 & 0.4 & 0.3 \\
     & 0.2 & 0.5 & 0.4 & 0.4 \\
     & 0.3 & 0.6 & 0.5 & 0.4 \\
     & 0.4 & 0.7 & 0.5 & 0.5 \\
     & 0.5 & 0.8 & 0.6 & 0.5 \\
     & 0.6 & - & 0.7 & 0.6 \\
     &0.7 &  - & 0.70 & 0.40 \\
     &0.8 &  - & 0.80 & 0.45 \\
     &0.9 &  - & 0.80 & 0.50 \\
     &1.0 &  - & - & 0.55 \\
\tableline
II & 0.1 &   0.5 & 0.4 & 0.4 \\
   & 0.2 &   0.6 & 0.4 & 0.4 \\
   & 0.3 &   0.6 & 0.5 & 0.5 \\
   & 0.4 &   0.7 & 0.5 & 0.5 \\
   & 0.5 &   0.7 & 0.6 & 0.5 \\
   & 0.6 &   0.8 & 0.6 & 0.6 \\
   &0.7 &  0.8 & 0.70 & 0.6 \\
   &0.8 &  0.9 & 0.80 & 0.70 \\
   &0.9 &  1.0 & 0.90 & 0.70 \\
   &1.0 &  - & - & 0.70 \\
\tableline
\tableline
\end{tabular}
\end{center}
\end{table}

\end{document}